\documentclass[fleqn,usenatbib]{mnras}

\usepackage{newtxtext,newtxmath}

\usepackage[T1]{fontenc}

\DeclareRobustCommand{\VAN}[3]{#2}
\let\VANthebibliography\thebibliography
\def\thebibliography{\DeclareRobustCommand{\VAN}[3]{##3}\VANthebibliography}

\usepackage{graphicx}	
\usepackage{amsmath}	
\usepackage{ae,aecompl}
\usepackage{lscape}
\newcommand{\orc}{\includegraphics[height=\fontcharht\font`A]{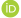}}
\newcommand{\orcid}[1]{\href{https://orcid.org/#1}{\orc}}

\def\nodata{ ~$\cdots$~ }

\newcommand{\ms}{m\,s$^{-1}$}
\newcommand{\masy}{mas\,y$^{-1}$}
\newcommand{\mpl}{\mbox{M$_{P}$}}
\newcommand{\rpl}{\mbox{R$_{P}$}}
\newcommand{\rhopl}{\mbox{$\rho_{P}$}}
\newcommand{\mstar}{\mbox{M$_{*}$}}
\newcommand{\rstar}{\mbox{R$_{*}$}}

\newcommand{\mjup}{\mbox{M$_{J}$}}
\newcommand{\rjup}{\mbox{R$_{J}$}}

\newcommand{\msun}{\mbox{M$_{\odot}$}}
\newcommand{\rsun}{\mbox{R$_{\odot}$}}
\newcommand{\gccc}{g\,cm$^{-3}$}

\newcommand{\teff}{$T_{\rm eff}$}
\newcommand{\feh}{\mbox{[Fe/H]}}

\newcommand{\logg}{$\log g$}
\newcommand{\tc}{$T_C$}
\newcommand{\rprs}{\mbox{R$_{P}/$R$_{*}$}}
\newcommand{\Astartic}{TIC-201177276}
\newcommand{\Astartoi}{TOI-2379}
\newcommand{\Atwomass}{2MASS J23522205-5302354}
\newcommand{\AGAIAid}{6521531466699512064}

\newcommand{\AgammaESP}{\mbox{$27718.9\pm5.1$}}
\newcommand{\AgammaPFS}{\mbox{$-816\pm54$}}
\newcommand{\AjitterESP}{\mbox{$0\pm17$}}
\newcommand{\AjitterPFS}{\mbox{$156\pm46$}}

\newcommand{\ARA}{\mbox{$23^{\rmn{h}} 53^{\rmn{m}} 22\fs09$}} 
\newcommand{\ADec}{\mbox{$-53\degr 02\arcmin 35\farcs 36$}} 
\newcommand{\ApropRA}{\mbox{$16.528\pm0.029$}} 
\newcommand{\ApropDec}{\mbox{$1.740\pm0.036$}} 

\newcommand{\Astarmasseccen}{\mbox{$0.645\pm0.033$}}
\newcommand{\Astarradiuseccen}{\mbox{$0.622\pm0.011$}}
\newcommand{\Astardensityeccen}{\mbox{$3.76\pm0.10$}}
\newcommand{\AteffODUSSEAS}{\mbox{$3664 \pm 66$}}
\newcommand{\Ateffeccen}{\mbox{$3707\pm58$}}
\newcommand{\AmetalODUSSEAS}{\mbox{$0.08 \pm 0.10$}} 
\newcommand{\Ametaleccen}{\mbox{$0.501\pm0.085$}} 
\newcommand{\Adist}{\mbox{$211.4\pm 1.1$}}
\newcommand{\Aplx}{\mbox{$4.705\pm0.024$}}

\newcommand{\Aloggeccen}{\mbox{$4.653 \pm 0.011$}}

\newcommand{\Aage}{$10.4_{-3.5}^{+5.1}$} 
\newcommand{\AVmag}{$15.340 \pm 0.010$}
\newcommand{\ABmag}{$17.01 \pm 0.13$}

\newcommand{\ATESSmag}{$13.6521 \pm 0.0073$}
\newcommand{\AGAIAmag}{$14.66010 \pm 0.00030$}
\newcommand{\AGAIABpmag}{$15.6701 \pm 0.0028$}
\newcommand{\AGAIARpmag}{$13.6697 \pm 0.0012$}
\newcommand{\AJmag}{$12.479 \pm 0.031$}
\newcommand{\AHmag}{$11.715 \pm 0.042$}
\newcommand{\AKmag}{$11.517 \pm 0.026$}
\newcommand{\AWmag}{$11.446 \pm 0.023$}
\newcommand{\AWWmag}{$11.471 \pm 0.021$}
\newcommand{\AWWWmag}{$11.34 \pm 0.13$}

\newcommand{\Aplanettoi}{TOI-2379\,b}
\newcommand{\Aperiod}{\mbox{$5.4693827 \pm 0.0000023$}}
\newcommand{\Aperiodshort}{\mbox{$5.469$}}
\newcommand{\Aduration}{\mbox{$1.867 \pm 0.012$}}%
\newcommand{\Atc}{\mbox{$2459401.56596 \pm 0.00016$}}
\newcommand{\Aecc}{\mbox{$0.3420\pm0.0039$}}%
\newcommand{\Aomega}{\mbox{$73.1\pm2.1$}}
\newcommand{\Amass}{\mbox{$5.76\pm0.20$}}%
\newcommand{\Aradius}{\mbox{$1.046\pm0.023$}}%
\newcommand{\Adensity}{\mbox{$6.21\pm0.29$}}%
\newcommand{\Arratio}{\mbox{$0.1729\pm0.0018$}}
\newcommand{\Aau}{\mbox{$0.05263\pm0.00091$}}%
\newcommand{\Aaoverr}{\mbox{$18.17\pm0.16$}}%

\newcommand{\Aimpact}{\mbox{$0.242_{-0.062}^{+0.043}$}}%
\newcommand{\Asemiamp}{\mbox{$1002.5\pm6.3$}}
\newcommand{\Ainc}{\mbox{$88.85_{-0.21}^{+0.30}$}}
\newcommand{\Amassratio}{\mbox{$0.85 \pm 0.05$}}
\newcommand{\Ateqave}{\mbox{$624\pm12$}}
\newcommand{\Ainsolave}{\mbox{$(3.45 \pm 0.22) \times 10^7$}}

\newcommand{\Bstartic}{TIC-382602147}
\newcommand{\Bstartoi}{TOI-2384}
\newcommand{\Btwomass}{2MASS J02243746-6459599}
\newcommand{\BGAIAid}{4699702272124241152}

\newcommand{\BgammaESP}{\mbox{$18059.3\pm8.3$}}
\newcommand{\BjitterESP}{\mbox{$16.3\pm8.5$}}

\newcommand{\BRA}{\mbox{$02^{\rmn{h}} 24^{\rmn{m}} 37\fs48$}} 
\newcommand{\BDec}{\mbox{$-65\degr 00\arcmin 00\farcs 7$}} 
\newcommand{\BpropRA}{\mbox{$8.58\pm0.11 $}} 
\newcommand{\BpropDec}{\mbox{$-52.648\pm0.095$}} 
\newcommand{\Bstarmass}{\mbox{$0.635\pm0.016$}}
\newcommand{\Bstarradius}{\mbox{$0.6113\pm0.0080$}}
\newcommand{\Bstardensity}{\mbox{$3.91\pm0.15$}}
\newcommand{\BteffODUSSEAS}{\mbox{$3609 \pm 67$}}
\newcommand{\Bteff}{\mbox{$3943\pm14$}}
\newcommand{\BmetalODUSSEAS}{\mbox{$0.17 \pm 0.10$}} 
\newcommand{\Bmetal}{\mbox{$0.332\pm0.063$}} 
\newcommand{\Bdist}{\mbox{$187.0 \pm 1.5$}}
\newcommand{\Bplx}{\mbox{$5.322 \pm 0.044$}}
\newcommand{\Blogg}{\mbox{$4.660 \pm 0.013$}}

\newcommand{\Bage}{$9.4\pm5.8$}
\newcommand{\BVmagorig}{$15.159 \pm 0.021$}
\newcommand{\BVmag}{$15.36 \pm 0.26$}
\newcommand{\BBmagorig}{$16.63 \pm 0.014$}
\newcommand{\BBmag}{$16.83 \pm 0.30$}

\newcommand{\BTESSmag}{$13.3147 \pm 0.0074$}
\newcommand{\BGAIAmag}{$14.3880 \pm 0.0029$}
\newcommand{\BGAIABpmag}{$15.3878 \pm 0.0037$}
\newcommand{\BGAIARpmag}{$13.3216 \pm 0.0043$}
\newcommand{\BJmagorig}{$11.997 \pm 0.026$}
\newcommand{\BJmag}{$12.27 \pm 0.10$}
\newcommand{\BHmagorig}{$11.308 \pm 0.024$}
\newcommand{\BHmag}{$11.489 \pm 0.098$}
\newcommand{\BKmagorig}{$11.119 \pm 0.023$}
\newcommand{\BKmag}{$11.341 \pm 0.098$}

\newcommand{\BWmag}{$11.22 \pm 0.11$}
\newcommand{\BWWmag}{$11.29 \pm 0.14$}
\newcommand{\BWWWmag}{$10.87 \pm 0.17$}
\newcommand{\BWmagorig}{$11.016 \pm 0.022$}
\newcommand{\BWWmagorig}{$11.036 \pm 0.020$}
\newcommand{\BWWWmagorig}{$10.604 \pm 0.067$}

\newcommand{\Bplanettoi}{TOI-2384\,b}
\newcommand{\Bperiod}{\mbox{$2.13570304\pm0.00000038$}}
\newcommand{\Bperiodshort}{\mbox{$2.136$}}
\newcommand{\Bduration}{\mbox{$1.891\pm0.012$}}%
\newcommand{\Btc}{\mbox{$2459591.99589\pm0.000117$}}
\newcommand{\Bmass}{\mbox{$1.966\pm0.059$}}%
\newcommand{\Bradius}{\mbox{$1.025\pm0.021$}}%
\newcommand{\Bdensity}{\mbox{$2.26\pm0.15$}}%
\newcommand{\Brratio}{\mbox{$0.1723\pm0.0024$}}
\newcommand{\Bau}{\mbox{$0.02793\pm0.00023$}}%
\newcommand{\Baoverr}{\mbox{$9.82\pm0.12$}}%

\newcommand{\Bimpact}{\mbox{$0.280_{-0.060}^{+0.046}$}}%
\newcommand{\Bsemiamp}{\mbox{$419\pm10$}}
\newcommand{\Binc}{\mbox{$88.37\pm0.34$}}
\newcommand{\Bmassratio}{\mbox{$0.30 \pm 0.01$}}
\newcommand{\Bteq}{\mbox{$889.7\pm5.0$}}
\newcommand{\Binsol}{\mbox{$(1.420\pm0.032) \times 10^8$}}

\title[\Aplanettoi\ \& \Bplanettoi]{\Aplanettoi\ and \Bplanettoi : two super-Jupiter mass planets transiting low-mass host stars\thanks{This paper includes data gathered with the 6.5 meter Magellan Telescopes located at Las Campanas Observatory, Chile.}}

\author[E. M. Bryant et al.]{
\parbox{\textwidth}{
Edward M. Bryant\orcid{0000-0001-7904-4441},$^{1}$\thanks{E-mail:\href{edward.bryant@ucl.ac.uk}{ edward.bryant@ucl.ac.uk}}
Daniel Bayliss\orcid{0000-0001-6023-1335},$^{2, 3}$
Joel D. Hartman\orcid{0000-0001-8732-6166},$^{4}$
Elyar Sedaghati\orcid{0000-0002-7444-5315},$^{5}$
Melissa J. Hobson\orcid{0000-0002-5945-7975},$^{6,7}$
Andr\'es Jord\'an\orcid{0000-0002-5389-3944},$^{7,8,9}$
Rafael Brahm\orcid{0000-0002-9158-7315},$^{7,8,9}$
Gaspar \'A. Bakos\orcid{0000-0001-7204-6727},$^{4}$
Jose Manuel Almenara,$^{10, 11}$ 
Khalid Barkaoui\orcid{0000-0003-1464-9276},$^{12,13,14}$ 
Xavier Bonfils,$^{10}$
Marion Cointepas,$^{10, 11}$ 
Karen A. Collins,$^{15}$ 
Georgina Dransfield,$^{16}$ 
Phil Evans\orcid{0000-0002-5674-2404},$^{17}$ 
Micha\"el Gillon,$^{12}$ 
Emmanu\"el Jehin,$^{18}$ 
Felipe Murgas\orcid{0000-0001-9087-1245},$^{14,19}$ 
Francisco J. Pozuelos,$^{20, 12}$ 
Richard P. Schwarz\orcid{0000-0001-8227-1020},$^{15}$ 
Mathilde Timmermans,$^{12}$ 
Cristilyn N. Watkins,$^{15}$ 
Ana\"{e}l W\"{u}nsche,$^{21}$
R. Paul Butler\orcid{0000-0003-1305-3761},$^{22}$ 
Jeffrey D. Crane\orcid{0000-0002-5226-787X},$^{23}$ 
Steve Shectman\orcid{0000-0002-8681-6136},$^{23}$ 
Johanna K. Teske\orcid{0000-0002-8681-6136}, $^{22}$ 
David Charbonneau,$^{15}$ 
Zahra Essack\orcid{0000-0002-2482-0180},$^{24}$ 
Jon M. Jenkins\orcid{0000-0002-4715-9460},$^{25}$ 
Hannah M. Lewis\orcid{0000-0002-7871-085X},$^{26}$ 
Sara Seager\orcid{0000-0002-6892-6948},$^{27,13,28}$ 
Eric B. Ting\orcid{0000-0002-8219-9505},$^{25}$ 
Joshua N. Winn\orcid{0000-0002-4265-047X}$^{4}$ 
}\\
\textit{Affiliations listed in Appendix~\ref{sec:affil}.}
}

\date{Accepted 2024 August 22. Received 2024 August 20; in original form 2024 February 16}

\pubyear{2024}

\begin{document}
\label{firstpage}
\pagerange{\pageref{firstpage}--\pageref{lastpage}}
\maketitle
\begin{abstract}
Short-period gas giant planets have been shown to be significantly rarer for host stars less massive than the Sun. We report the discovery of two transiting giant planets - \Aplanettoi\ and \Bplanettoi\ - with low-mass (early M) host stars. Both planets were detected using TESS photometry and for both the transit signal was validated using ground based photometric facilities. We confirm the planetary nature of these companions and measure their masses using radial velocity observations. We find that \Aplanettoi\ has an orbital period of \Aperiodshort\,d and a mass and radius of \Amass\,\mjup\ and \Aradius\,\rjup\ and \Bplanettoi\ has an orbital period of \Bperiodshort\,d and a mass and radius of \Bmass\,\mjup\ and \Bradius\,\rjup. \Aplanettoi\ and \Bplanettoi\ have the highest and third highest planet-to-star mass ratios respectively out of all transiting exoplanets with a low-mass host star, placing them uniquely among the population of known exoplanets and making them highly important pieces of the puzzle for understanding the extremes of giant planet formation.
\end{abstract}

\section{Introduction}

Close-in gas giant planets ($P\leq10$\,d; \rpl$\geq0.6$\rjup), known as hot Jupiters, dominated early transiting exoplanet discoveries \citep[e.g.][]{charbonneau2000hd209458,alonso2004tres1,bakos2007hatp1,mccullough2006xo1,colliercameron2007wasp1}. Despite this, and the relative ease of finding a hot Jupiter compared to a smaller planet, few hot Jupiters have been discovered orbiting M-dwarf host stars. This is in part due to the fact that these objects are significantly rarer around low-mass host stars \citep{bryant2023lmstargiantplanetoccrates, gan2023earlymgiantplanetoccrates}, but also due to the fact that obtaining the precise radial velocity measurements necessary to measure the mass of and confirm these planets is much harder for low-mass stars. Firstly this is due to the fact that low-mass stars have a low luminosity with this intrinsic faintness reducing the radial velocity precision achievable. Moreover M-dwarf spectra display a large amount of molecular lines, compared to the sharp metal lines present in the spectra of solar-like stars, further increasing the difficulty of achieving the high-precision radial velocity measurements required.

The rarity of these planets has been linked to the fact that lower mass stars are less likely to form giant planets through the core-accretion planet model \citep[e.g.][]{laughlin2004coreaccretion, kennedy2008formation, burn2021ngppslowmassstars}. This is primarily due to the lower surface density of the protoplanetary disks but the slower Keplerian timescales around these low-mass stars also slows down and inhibits planet formation as well \citep[e.g.][]{laughlin2004coreaccretion, idalin2005planetformation}. For early-M-type stars ($0.5\msun \lesssim \mstar \lesssim 0.65\msun$) giant planet formation through core-accretion is expected to be less likely than for a solar-mass star but not impossible \citep{burn2021ngppslowmassstars}. A number of discoveries of such planets over recent years speak to the potential, albeit rare, to form these planets \citep[e.g.][]{bayliss2018ngts1,bakos2020hats71,jordan2022hats74_77,kanodia2022toi3757,hartman:2023,canas2023toi3984toi5293} . For lower-mass stars ($\mstar \leq 0.4\msun$) the predictions are that giant planet formation becomes significantly more difficult, and yet recent discoveries of giant planets with such low-mass stars \citep[e.g.][]{kanodia2023toi5205,Hobson2023toi3235,almenara2023toi4860} have shown formation of these exotic systems to still be possible. While core-accretion struggles to form giant planets with lower mass host stars, there is an alternative pathway through which these systems could form: disk instability \citep[e.g.][]{boss1997diskinstability}. In fact, simulations have shown that giant planets can readily form through disk instability around M-dwarf stars \citep[e.g.][]{boss2006mdwarfdiskinstability, mercerstamatellos2020diskinstability, boss2023gravinstabilityMdwarfs}, with this pathway resulting in very massive ($\gtrsim 2.0$\,\mjup) planets \citep{mercerstamatellos2020diskinstability}. By revealing and studying the growing population of giant planets with low-mass host stars, we can better understand how the mass of the host star influences the formation of giant planets.

In addition to the stellar mass, the metallicity of the host star can also play a major role in the formation of giant planets. Specifically, a correlation between the occurrence rate of giant planets orbiting FGK stars and the host star metallicity has been observed and explained as a consequence of core-accretion planet formation \citep{fischer2005pmc,johnson2010gplmetallicity,wang2018PMCcoreaccretion,osborn2020pmc}. Similar trends showing enhanced host stars metallicities have also been shown for low-mass host stars of giant planets \citep{johnsonapps2009mdwarfPMC,rojasayala2010mdwarfPMC}. Theoretical studies have shown the core-accretion paradigm to form giant planets more easily in metal-rich protoplanetary disks \citep[e.g.][]{idalin2004metallicityformation, emsenhuber2021ngpps2} and the metallicity of a protoplanetary disk is believed to be inherited from the same primordial cloud as the host star, therefore resulting in more metal-rich disks around high metallicity stars \citep[e.g.][]{fischer2005pmc}. 

To add to the sample of planets that can be used to shed light on the formation history of these exotic planets and understand the roles played by these different formation effects for low-mass stars we present the discovery of \Aplanettoi\ and \Bplanettoi\ -- two super-Jupiter mass giant planets transiting metal-rich ($\feh > 0.35$dex) early M-dwarf stars. These two objects were discovered as a result of the all-sky coverage of the Transiting Exoplanet Survey Satellite \citep[TESS;][]{ricker2015tess}, and confirmed with ground-based photometric and spectroscopic follow-up. Their high masses, along with the low masses of their host stars, make them important additions to the known population of exoplanets, particularly from the viewpoint of understanding giant planet formation. Our discovery and follow-up photometric observations are presented in Section~\ref{sec:phot} and we use these to perform an initial analysis to confirm the true sources of the observed transiting signals which we present in Section~\ref{sec:nearby_blends}. Our spectroscopic observations are discussed in Section~\ref{sec:spec}. We discuss the analysis performed to derive the stellar and planetary characteristics in Section~\ref{sec:analysis} and we place these two exotic planets in context of the overall exoplanet population in Section~\ref{sec:discuss}.

\section{Photometric Observations}\label{sec:phot}
\subsection{TESS Photometry}\label{sub:tessphot}
The Transiting Exoplanet Survey Satellite \citep[TESS; ][]{ricker2015tess} is a NASA space-based exoplanet hunting facility that has been continuously observing the sky searching for transiting exoplanets since July 2018. The wide-field, high precision observations of TESS along with the fact TESS surveys the full sky on a timescale of two years make it an ideal data source for the discovery of rare astrophysical objects, such as giant planets orbiting low-mass stars. In fact, of the 21 known giant planets transiting low-mass stars ($\mstar \leq 0.65\,\msun$) TESS photometry was used to discover 14 \citep[e.g.][]{gan2022toi530, canas2022toi3714toi3629, kanodia2023toi5205} and was involved in the confirmation of 3 more \citep{bakos2020hats71, jordan2022hats74_77}.  

During the first year of TESS observations \Astartoi\ (\Astartic) was observed in Sectors~1~and~2 (25 July 2018 to 20 September 2018) and \Bstartoi\ (\Bstartic) was observed in Sectors~1,~2,~and~3 (25 July 2018 to 18 October 2018). Across these sectors, both stars were observed in the Full-Frame-Images (FFIs), with the FFIs in Year 1 of TESS being supplied to the community at a cadence of 30\,minutes. Light curves for both these stars were generated using a custom difference imaging analysis pipeline \citep[DIA\footnote{Code available at \url{https://github.com/ryanoelkers/DIA}}; see][]{oelkersstassun2018dia} and made available to the public through the \textit{Filtergraph} platform\footnote{\url{https://filtergraph.com/tess_ffi}}. We performed an independent search for transiting giant planets with low-mass host stars using the DIA light curves for the first five TESS sectors. This transit search was performed using the \textsc{astropy} implementation of the Box-fitting Least Squares algorithm \citep[BLS;][]{kovacs2002bls,astropy3_2022}. From this BLS search, we identified both \Astartoi\ and \Bstartoi\ as potential hosts of transiting giant planets. In TESS light curve of \Astartoi\ we detected a transit signal with a depth of 3\% and a period of 5.463\,d and in the TESS light curve of \Bstartoi\ we detected a transit signal with a depth of 2\% and a period of 2.136\,d. 

After the initial identification of these two systems as transiting giant planet candidates, we successfully proposed they be included as 2\,minute cadence targets in Year 3 of TESS observations (GI program G03129; PI Bryant). In Year 3 of TESS operations \Astartoi\ and \Bstartoi\ were both observed at 2\,minute cadence in Sectors~28~and~29 (30 July 2020 to 22 September 2020). The image data were processed by the Science Processing Operations Center (SPOC) at NASA Ames Research Center \citep{jenkins2016spoc} to extract photometry from these targets. For this work we use the \textsc{PDC\_SAP} light curves produced by the SPOC \citep{Stumpe2012pdcsap,Stumpe2014pdcsap,smith2012pdcsap}. The transit signature of \Aplanettoi\ was detected by the SPOC in the transit search of Sector~29 with a noise-compensating matched filter \citep{jenkins2002transitdetection,jenkins2010keplerTPS,jenkins2020TPSkdph}. A period of 5.470\,days was identified and the transit signature passed all the diagnostic tests presented in the resulting Data Validation report \citep{Twicken:DVdiagnostics2018}. A difference image centroiding test was performed and the source of the transit signature was located to within $2.622\pm2.57$\,\arcsec\ of \Astartoi. A subsequent analysis performed on data from both sectors 28 and 29 further constrained this location to within $0.58\pm3.2$\,\arcsec. \Aplanettoi\ was alerted as a TESS Object of Interest by the TESS Science Office on 6 November 2020 \citep{guerrero2021toicatalogue}. The transit signature of \Bplanettoi\ was detected in the transit search of Sector~28 at a period of 2.136\,days. The difference image centroiding test located the source of the transit signal to within $0.553\pm2.51$\,\arcsec\ of \Bstartoi. \Bplanettoi\ was also alerted as a TESS Object of Interest by the TESS Science Office on 6 November 2020 \citep{guerrero2021toicatalogue}. 

\begin{figure*}
    \centering
    \includegraphics[width=1.0\textwidth]{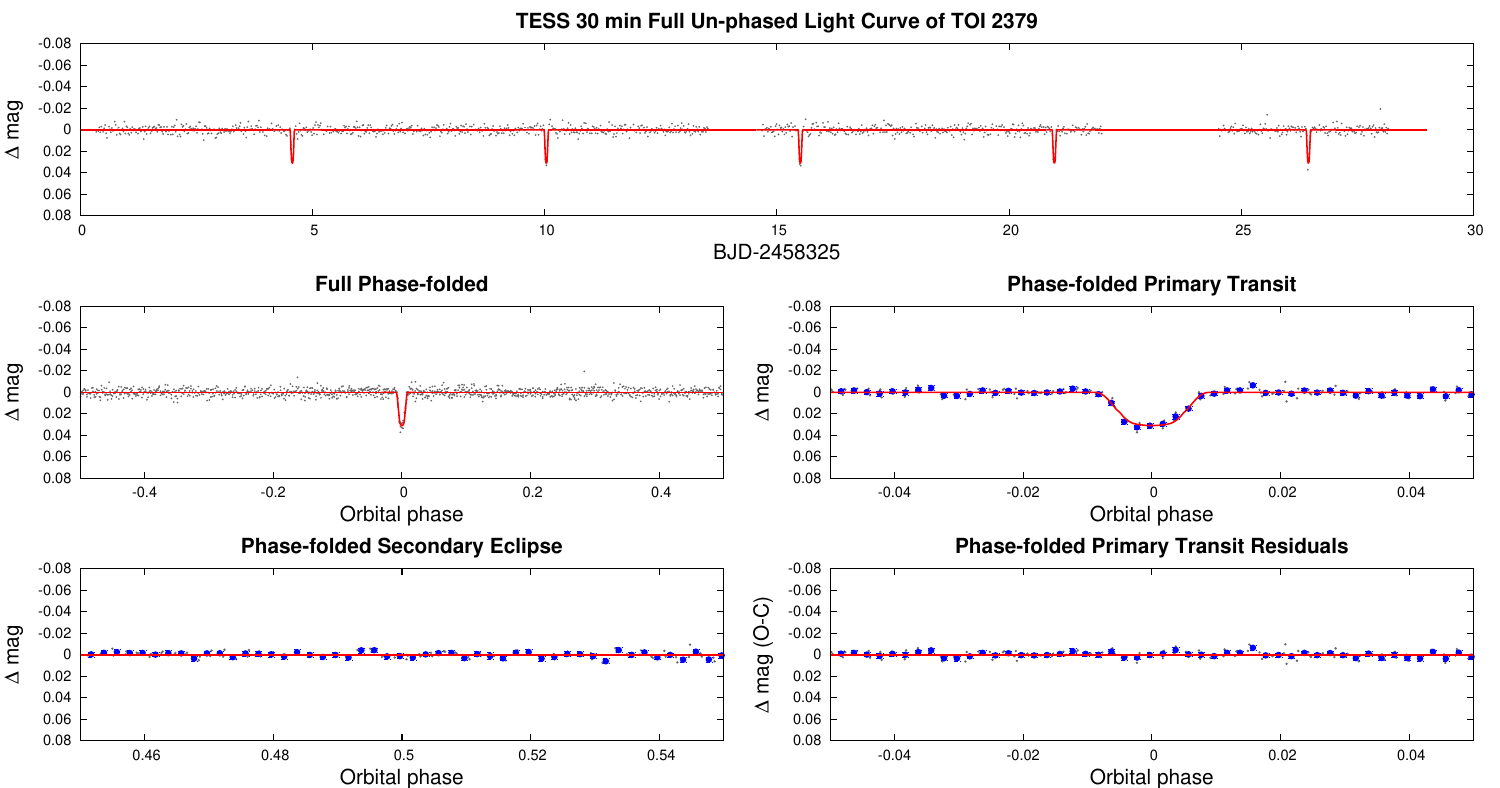}
    \caption{TESS photometry at a cadence of 30\,minutes \Astartoi. \textbf{Top:} the 30\,minute cadence TESS light curve as a function time. \textbf{Middle:} the TESS photometry phase-folded on the best fit ephemerides showing both the full light curve (\textit{left}) as well as a zoomed-in view around the transit event (\textit{right}). \textbf{Bottom left:} the TESS photometry phase-folded and zoomed-in on the location of the secondary eclipse. \textbf{Bottom right:} the residuals to the best-fit model (Section~\ref{sec:analysis}) zoomed-in around the transit event. For all the panels the grey points show the unbinned data, the larger blue circles show the data binned to a timescale of 0.002 in phase, and solid red line shows the best-fit model to the photometry.}
    \label{fig:toi2379_tess}
\end{figure*}

\begin{figure*}
    \centering
    \includegraphics[width=1.0\textwidth]{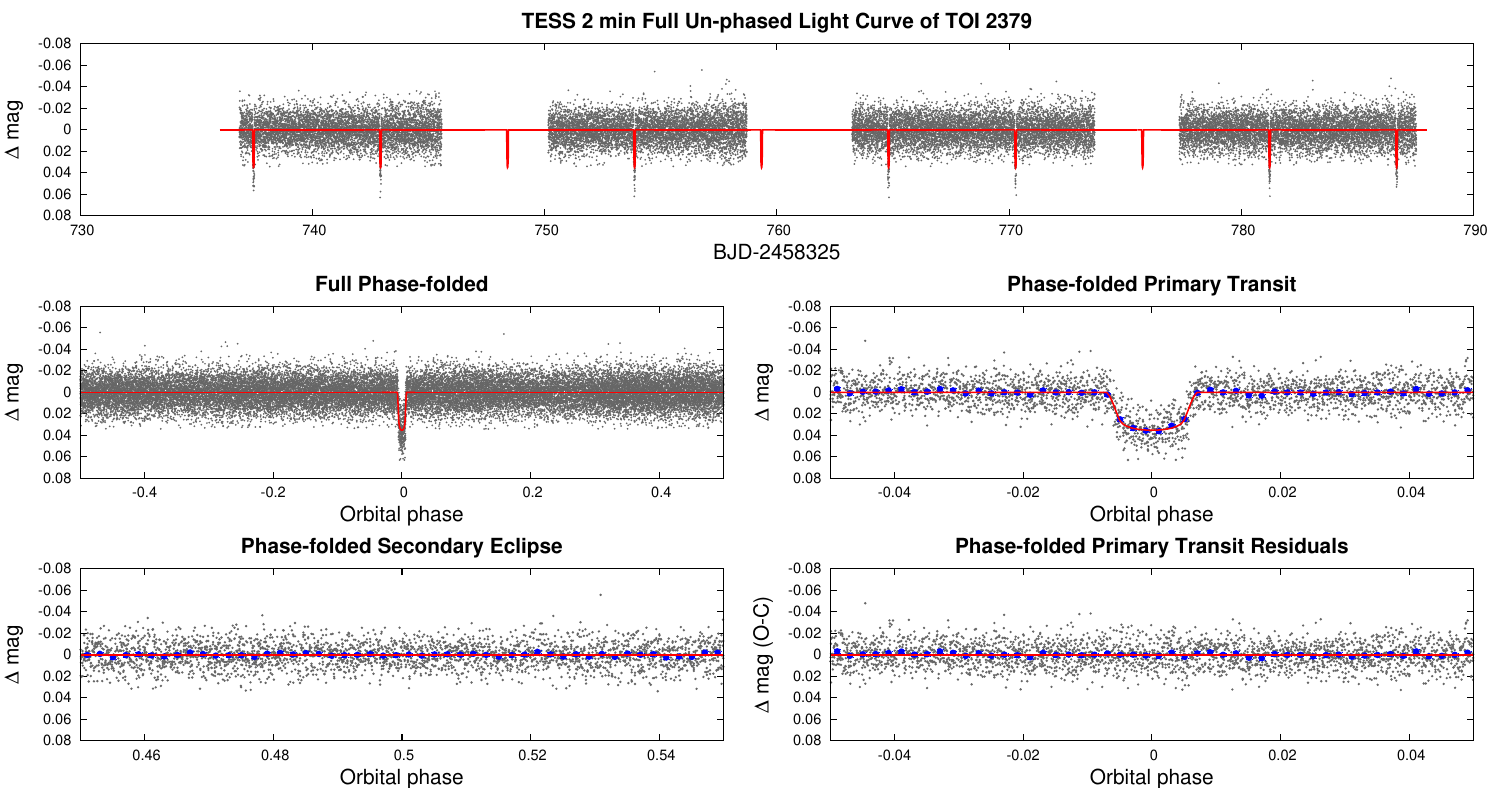}
    \caption{TESS photometry at a cadence of 2\,minutes for \Astartoi. The panels and layout are the same as Figure~\ref{fig:toi2379_tess}.}
    \label{fig:toi2379_tess2}
\end{figure*}

\begin{figure*}
    \centering
    \includegraphics[width=1.0\textwidth]{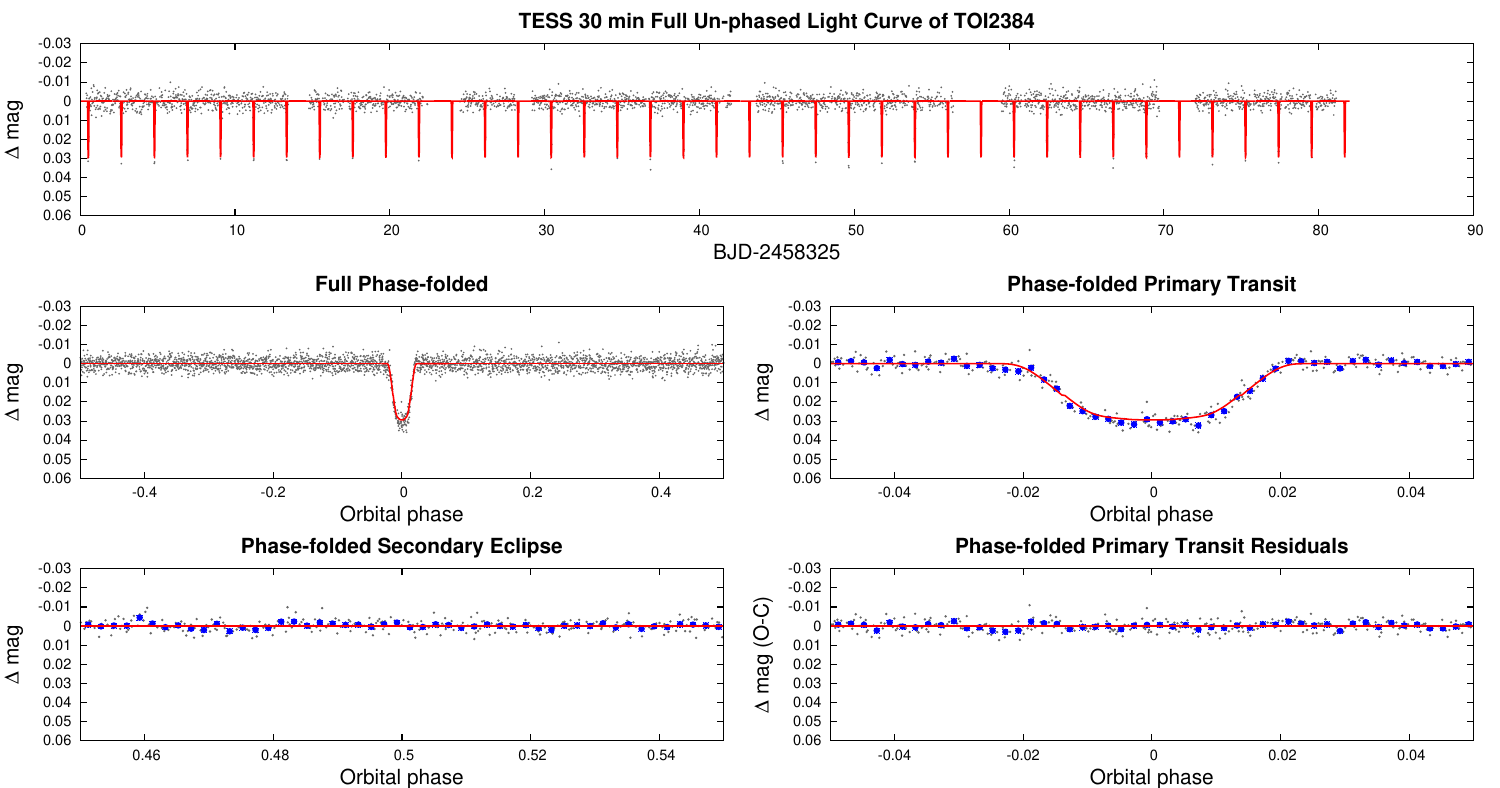}
    \caption{TESS photometry at a cadence of 30\,minutes for \Bstartoi. The panels and layout are the same as Figure~\ref{fig:toi2379_tess}.}
    \label{fig:toi2384_tess}
\end{figure*}

\begin{figure*}
    \centering
    \includegraphics[width=1.0\textwidth]{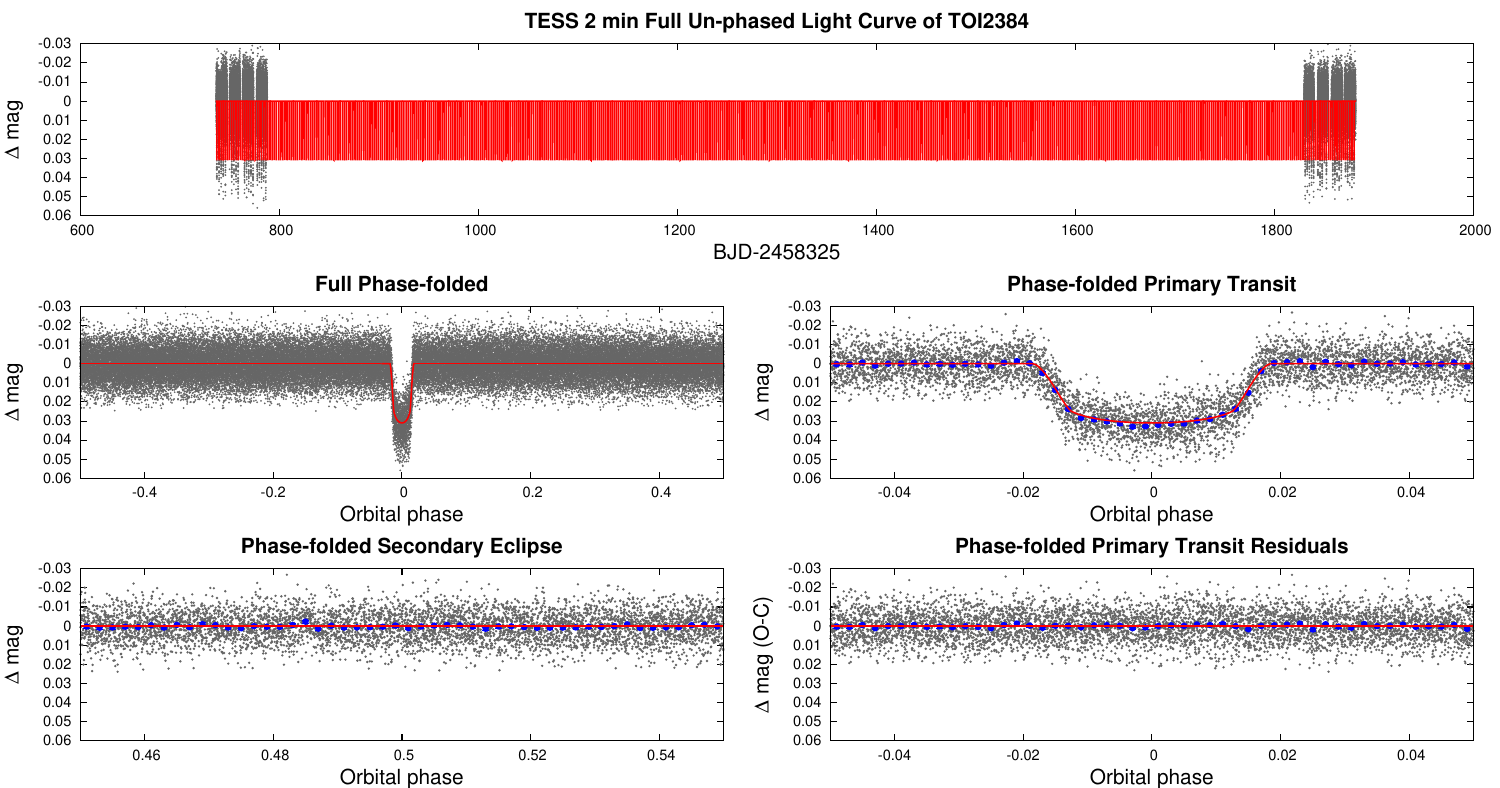}
    \caption{TESS photometry at a cadence of 2\,minutes for \Bstartoi. The panels and layout are the same as Figure~\ref{fig:toi2379_tess2}.}
    \label{fig:toi2384_tess2}
\end{figure*}

TESS photometry for these two objects is shown in Figures~\ref{fig:toi2379_tess},~\ref{fig:toi2379_tess2},~\ref{fig:toi2384_tess}~and~\ref{fig:toi2384_tess2}. For the analysis presented in Section~\ref{sec:analysis}, for \Astartoi\ we use an image-subtraction-based light curve extracted from the Sector~1 30\,minute cadence FFI observations following the methods of \citet{bouma2019cdips} and the SPOC 2\,minute cadence light curves for Sectors~28~and~29. For \Bstartoi\ we use 30\,minute cadence light curves extracted from the FFIs using the Quick Look Pipeline \citep[QLP;][]{huang2020qlp,kunimoto2021qlp} for Sectors~1,~2,~and~3 and the SPOC 2\,minute cadence light curves for Sectors~28~and~29. We accessed these light curves from the Mikulski Archive for Space Telescopes\footnote{\url{https://archive.stsci.edu/}} (MAST) and the choice of light curves used was motivated by the data publicly available from MAST at the time the analysis was performed.

\subsection{Follow-Up Photometry}\label{sub:followupphot}
\begin{figure}
    \centering
    \includegraphics[width=\columnwidth]{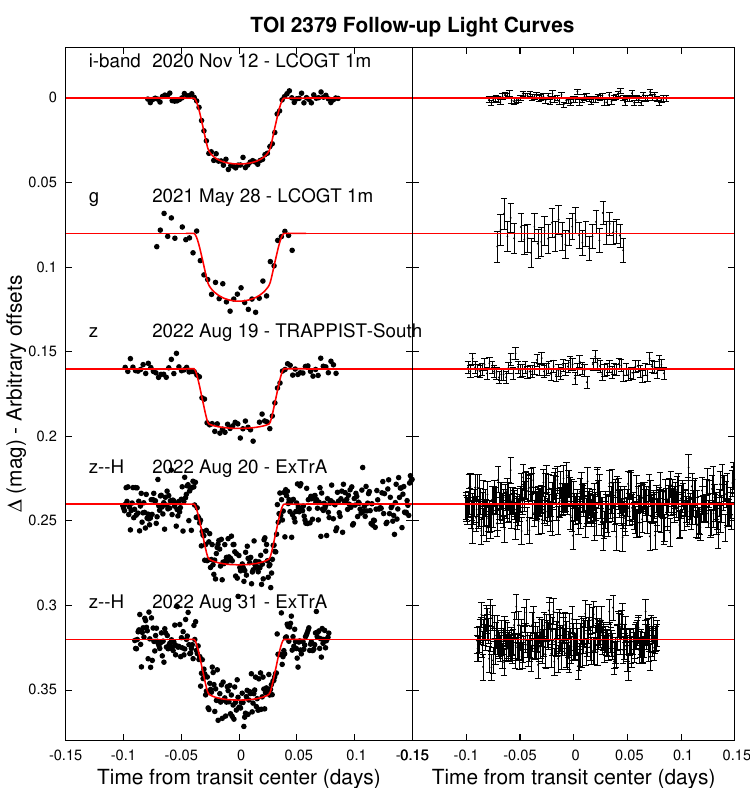}
    \caption{Ground based follow-up photometry for \Astartoi. The left panel shows each of the transit light curves obtained, plotted as a function of the time from the transit mid-point. Each light curve is labelled with the date on which the transit was observed along with the telescope and filter used to perform the observations. The right panel shows the residuals to the model fit for each light curve along with the photometric uncertainties for the light curves. For both panels, the points show the unbinned photometry data and the red solid lines give the best-fit models to the transit light curves. }
    \label{fig:toi2379_followupphot}
\end{figure}

\begin{figure}
    \centering
    \includegraphics[width=\columnwidth]{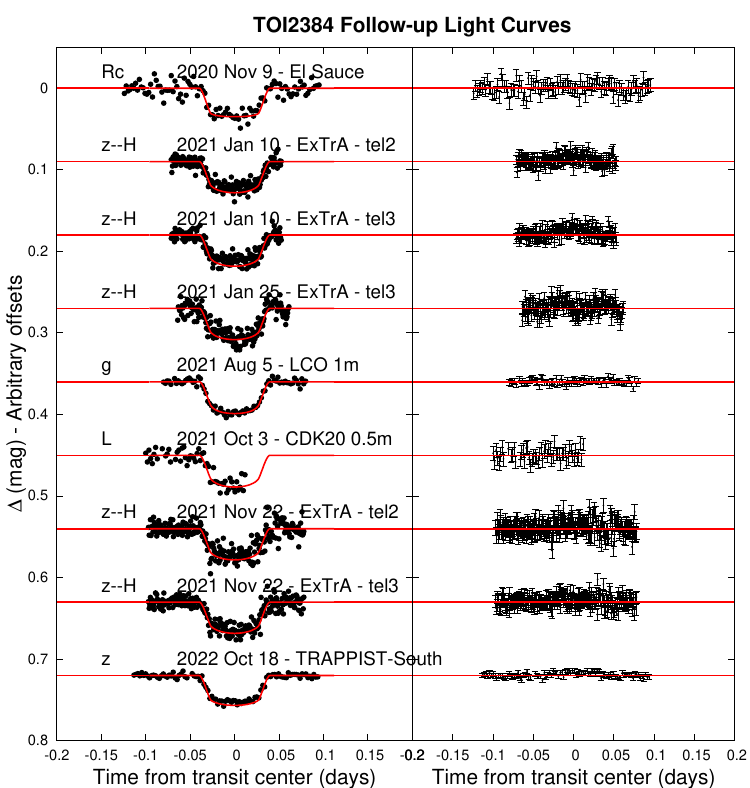}
    \caption{Ground based follow-up photometry for \Bstartoi. The panels and layout are the same as Figure~\ref{fig:toi2379_followupphot}. }
    \label{fig:toi2384_followupphot}
\end{figure}

The \textit{TESS} pixel scale is $\sim 21\arcsec$/pixel and photometric apertures typically extend out to roughly 1\arcmin, generally causing multiple stars to blend in the \textit{TESS} aperture. We conducted ground-based light curve follow-up observations of the field around both target stars as part of the \textit{TESS} Follow-up Observing Program\footnote{https://tess.mit.edu/followup} Sub Group 1 \citep[TFOP;][]{collins2019sg1}. The goals of these  observations were to determine the true source of the \Astartoi\ and \Bstartoi\ transit signals in the \textit{TESS} data, improve the transit ephemerides, and demonstrate consistent depths across multiple optical bands. We used the {\tt TESS Transit Finder}, which is a customized version of the {\tt Tapir} software package \citep{jensen2013}, to schedule our transit observations. Here we provide details on the follow-up light curves obtained and included in  the analysis for this work. The follow-up photometry for \Astartoi\ is shown in Figure~\ref{fig:toi2379_followupphot} and the follow-up photometry for \Bstartoi\ is shown in Figure~\ref{fig:toi2384_followupphot}.

\subsubsection{LCOGT}\label{phot:lco}
The Las Cumbres Observatory Global Telescope network \citep[LCOGT,][]{brown2013lcogt} is a globally distributed network of 1.0\,m telescopes. The telescopes are equipped with $4096\times4096$ SINISTRO cameras having an image scale of 0.389\arcsec/pixel, resulting in a $26\arcmin\times26\arcmin$ field of view. We observed two full transit windows of \Aplanettoi\ on 2020 November 12 and 2021 May 28 in Sloan $i'$ and Sloan $g'$ bands, respectively, at the LCOGT nodes at South Africa Astronomical Observatory (SAAO) and Siding Spring Observatory (SSO). We also observed one full transit window of \Bplanettoi\ on 2021 August 6 in Sloan $g'$ band using the LCOGT node at Cerro Tololo Inter-American Observatory (CTIO). The images were calibrated by the standard LCOGT {\tt BANZAI} pipeline \citep{McCully:2018} and differential photometric data were extracted using {\tt AstroImageJ} \citep{Collins:2017}. For \Astartoi, we used circular photometric apertures with radius $5\farcs8$. The target star apertures excluded all of the flux from the nearest known neighbor in the \textit{Gaia} DR3 catalog \citep[\textit{Gaia} DR3 6521531466700057856][]{GAIADR32021}, which is $\sim36\arcsec$ northwest of \Astartoi. For \Bstartoi, we used a circular photometric aperture with radius $4\farcs7$. The target star aperture is fully contaminated with the nearest known neighbor in the \textit{Gaia} DR3 catalog (\textit{Gaia} DR3 4699702272123475328), which is $\sim0\farcs9$ north of \Bstartoi. All three light curves are included in the global modelling described in Section~\ref{sec:analysis}. 

\subsubsection{TRAPPIST-South}\label{phot:trappist}

TRAPPIST-South \citep{jehin:2011,Gillon2011} is a 0.6\,m Ritchey-Chretien robotic telescope at La Silla Observatory in Chile, equipped with a 2K$\times$2K back-illuminated CCD camera with a pixel scale of 0.65\arcsec/pixel, resulting a field of view of $22\arcmin\times22\arcmin$. We observed a full transit of \Aplanettoi\ on 2022 August 19 and a full transit of \Bplanettoi\ on 2022 October 18. Both transits were observed in the Sloan-$z'$ filter with an exposure time of 180\,s. 
During the observations of both transits, the telescope underwent a meridian flip at BJD = 2459811.7816 for \Aplanettoi\ and at BJD = 2459871.72236 for \Bplanettoi.
Data reduction and photometric measurements were performed using the {\tt PROSE}\footnote{\textit{PROSE}: \url{https://github.com/lgrcia/prose}} pipeline \citep{garcia2021}.

\subsubsection{El Sauce}\label{phot:elsauce}
We observed a full transit of \Bplanettoi\ in Johnson-Cousins Rc-band on 2020 November 9 using the Evans 0.36\,m CDK14 telescope at El Sauce Observatory in Coquimbo Province, Chile. The telescope was equipped with a STT 1603-3 CCD camera with 1536$\times$1024 pixels binned 2$\times$2 in-camera resulting in an image scale of 1.47\arcsec/pixel. The photometric data was obtained from 106$\times$180\,s exposures, after standard calibration, using a circular 5.9\arcsec aperture in AstroImageJ \citep{Collins:2017}. The measurement aperture was fully contaminated with the nearest known \textit{Gaia} DR3 catalog neighbor (\textit{Gaia} DR3 4699702272123475328), which is 0.9\arcsec north of \Bstartoi\ and $\Delta T_{\rm{mag}}$ = 3.64.

A partial transit ingress of \Bstartoi\ was observed on 2021 October 3 using a 0.5\,m CDK20 telescope which is situated at El Sauce Observatory but is controlled remotely from the Observatory of Baronnies Proven\c{c}ales (OBP) in France. The OBP is a private observatory doing outreach, courses, training and research, addressed to all public and amateur and professional astronomers, which is part of the list of protected observatories for light pollution in France, in the regional park of Baronnies. The CDK20 telescope used is on a paramount equatorial mount and is equipped with a Moravian G4 16K CCD camera. The images were taken with 1$\times$1 binning and photometry was performed with an aperture of 9 pixels with the FWHM of the target estimated to be 2.3\arcsec\ with a pixel scale of 0.5255\arcsec/pixel. The analysis was performed using the Muniwin program from the photometry software package C-Munipack\footnote{\url{https://c-munipack.sourceforge.net/}} \citep{hroch2014munipack}.

\subsubsection{ExTrA}\label{phot:extra}
ExTrA \citep{Bon2015} is a low-resolution near-infrared (0.85 to 1.55~$\mu$m) multi-object spectrograph fed by three 60-cm telescopes located at La Silla Observatory in Chile. \Astartoi\ was observed on 2022 August 20 and 31. \Bstartoi\ was observed on 2021 January 10 and 25 and November 22. We used 8$\arcsec$ aperture fibers and the lowest-resolution mode ($R$$\sim$20) of the spectrograph, with an exposure time of 60~seconds. Five fibers are positioned in the focal plane of each telescope to select light from the target and four comparison stars. We chose comparison stars with 2MASS $J$-magnitudes \citep{2MASS} and effective temperatures \citep{GAIA_DR2} similar to the target. The resulting ExTrA data were analyzed using custom data reduction software.

\section{Nearby Blend Analysis}\label{sec:nearby_blends}
Due to the large pixels of the TESS images (21\,\arcsec~per~pixel) the observations can often suffer from blending from nearby stars. From the follow-up photometry presented in Section~\ref{sub:followupphot}, and in particular the LCOGT data, we can already rule out any known neighbouring \textit{Gaia} DR3 stars as the source of the transit signal seen for \Astartoi, and we can rule out all except the closest neighbour for \Bstartoi. We now perform some initial analyses to rule out this close neighbour as the true source of the transit signal observed in the TESS data and ground-based photometry for \Bstartoi\ as well as to check for any evidence either transit signal could be the result of a background blended source.

\begin{figure*}
    \centering
    \includegraphics[width=0.48\linewidth]{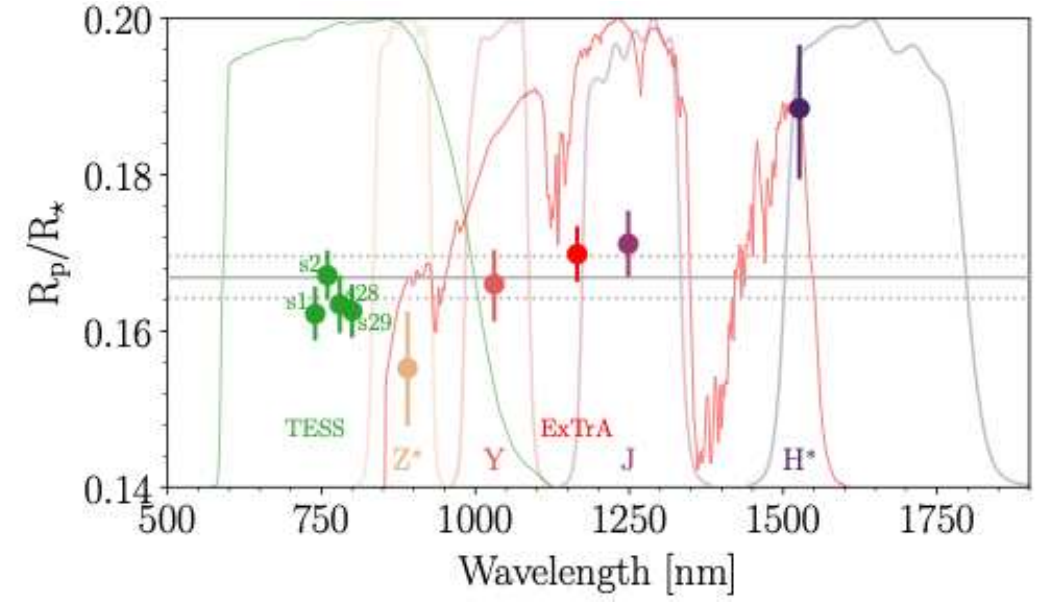}
    \hfill
    \includegraphics[width=0.48\linewidth]{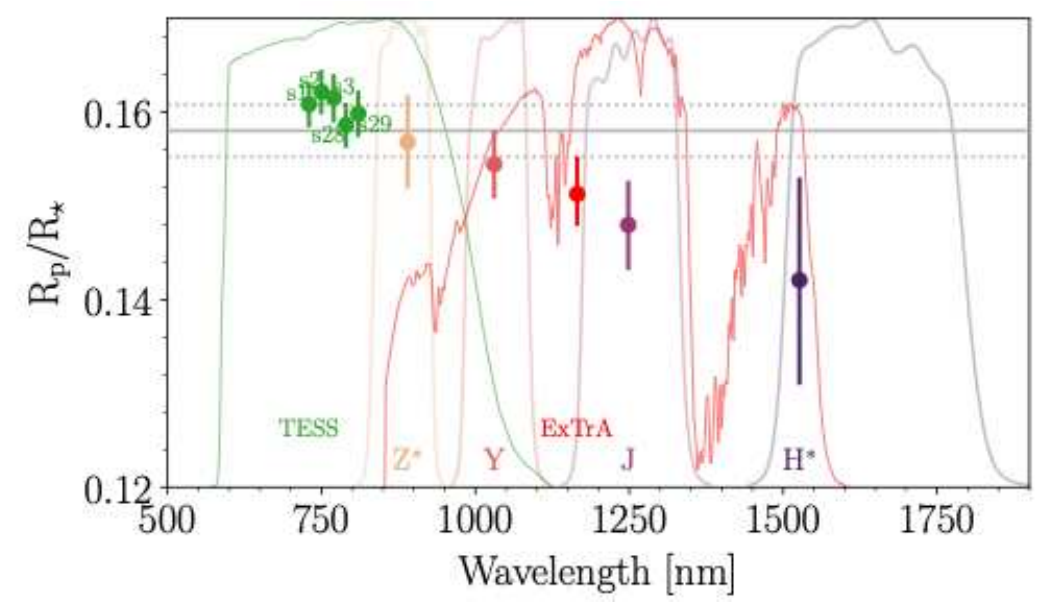}
    \caption{Chromaticity analysis results comparing the transit depths for \Aplanettoi\ (left panel) and \Bplanettoi\ (right panel) observed by TESS and ExTrA in the Z$^*$, Y, J, and H$^*$ bands. The circles and errorbars give the measured \rprs\ value for each filter, and each corresponding filter is labelled and plotted.}
    \label{fig:extra_chromaticity}
\end{figure*}

Firstly, the ExTrA spectrophotometric observations are used to generate four light curves in the UKIRT-WFCAM filters -- Y, J, and truncated versions of the Z and H bands, Z$^*$ and H$^*$ (see Figure~\ref{fig:extra_chromaticity}). Using these additional light curves we can perform a preliminary analysis to investigate whether the transit signals are observed to vary significantly with the wavelength of the observations. Such a chromatic dependence can be evidence that the transit signal is the result of a, possibly blended, stellar eclipsing binary. For more info on the analysis performed see Section 4.1 in \citet{almenara2023toi4860}, but in short the four ExTrA light curves were fit along with the TESS data with all planetary parameters except for the radius ratio \rprs\ held common between all data sets. 

The results of this analysis are plotted in Figure~\ref{fig:extra_chromaticity}. For \Aplanettoi\ we find consistent \rprs\ values across all filters except for H$^*$. However, the observations in this filter are low signal-to-noise, and so we do not find this conclusive evidence in favour of a blended scenario. Therefore, we conclude that the ExTrA multi-wavelength observations support the scenario of a single star with a transiting planet for the \Astartoi\ system. For \Bplanettoi\ we see a decrease in \rprs\ with increasing wavelength of the observations. \Bstartoi\ has a close neighbour which is both at the same distance as \Bstartoi\ and also significantly fainter \citep{gaiadr32021targetcatalogue} so is likely redder (see also Section~\ref{sec:analysis}). Therefore, the contribution of this neighbour star to the total flux observed within the ExTrA photometric aperture increases with increasing wavelength and so the observed decrease in \rprs\ with wavelength is most likely a result of this increased dilution. As such, the ExTrA observations again provide us with further evidence that transit event observed in the TESS light curve of \Bstartoi\ is on target.

Second, we employ the \textsc{TESSPositionalProbability} tool \citep{hadjigeorghiou2023tessposprob} to derive a probabilistic estimate of which star is the true source of the transit signals. \textsc{TESSPositionalProbability} uses the observed photometric centroid shift during the TESS transit events to estimate the likelihood that any of the nearby stellar sources identified by \textit{Gaia} could be the true source of the signal, and has been demonstrated to be highly accurate at determining the true source of TESS transit signals \citep{hadjigeorghiou2023tessposprob}. \textsc{TESSPositionalProbability} is designed to use the TESS Full-Frame-Image light curves which have been produced by the TESS-SPOC pipeline \citep{caldwell2020tessspoc} and made available as a MAST High-Level-Science-Product\footnote{\url{https://archive.stsci.edu/hlsp/tess-spoc}}. For both our targets, TESS-SPOC FFI light curves are available for Sectors~28~and~29, which we use for this analysis. From our \textsc{TESSPositionalProbability} analysis for \Astartoi\ we find that no nearby \textit{Gaia} stars can be the true source of the signal. For \Bstartoi\ we find that a probability of 99.7\% that \Bstartoi\ is the true source of the observed transit signal. The closest neighbour, which is heavily disfavoured as the true source based on our chromaticity analysis, is the only other star reported by \textsc{TESSPositionalProbability} as possible of being the source of the transit signal.

Combining these independent analyses with our follow-up photometry, we are confident that both \Astartoi\ and \Bstartoi\ are the true sources of the transiting signals.

\section{Spectroscopic Observations}\label{sec:spec}
\subsection{ESPRESSO}\label{spec:espresso}
To confirm the planetary nature of the transiting candidates we obtain spectroscopic observations using the ESPRESSO high resolution echelle spectrograph \citep{pepe2021espresso}, at the 8\,m  VLT facility at Paranal Observatory, Chile.  ESPRESSO first light was achieved in November 2017 and the commissioning was completed in July 2019. Since then ESPRESSO has successfully been employed in the follow-up of a variety of TOIs, particularly those with low-mass host stars \citep[e.g.][]{castrogonzalez2023toi244espresso,vaneylen2021toi270espresso,Hobson2023toi3235}. 

The ESPRESSO spectroscopic observations allow us to monitor the radial velocity variation of the stars, to determine the masses of the transiting companions. Both \Astartoi\ and \Bstartoi\ were observed as a part of program 108.22B4.001 (PI Jordan). We obtained seven observations of \Astartoi\ between 2021 October 9 and 2021 November 9 and seven observations of \Bstartoi\ between 2021 November 9 and 2021 December 4. We used an exposure time of 2400\,s for \Astartoi\ and 1800\,s for \Bstartoi. We reduced the spectra using the ESPRESSO DRS pipeline \citep[v2.3.5][]{sosnowska2015espressoDRS,modigliani2020espressoDRS} implemented in the EsoReflex environment \citep{freudling2013esoreflex}. The radial velocity and bisector spans for both objects are listed in Tables~\ref{tab:toi2379_rv}~and~\ref{tab:toi2384_rv} and presented in Figures~\ref{fig:toi2379_rv_SED}~and~\ref{fig:toi2384_rv_SED}.

\subsection{PFS}\label{PFS}\label{spec:pfs}

\begin{figure*}
    {
    \centering
    \includegraphics[width=0.49\linewidth]{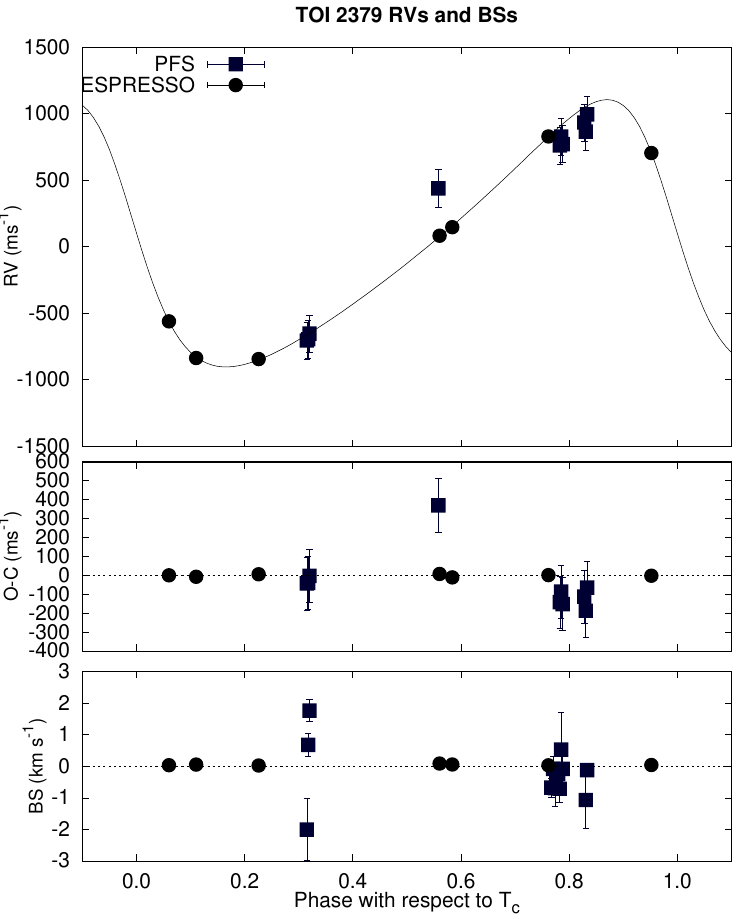}
    \hfill
    \includegraphics[width=0.49\linewidth]{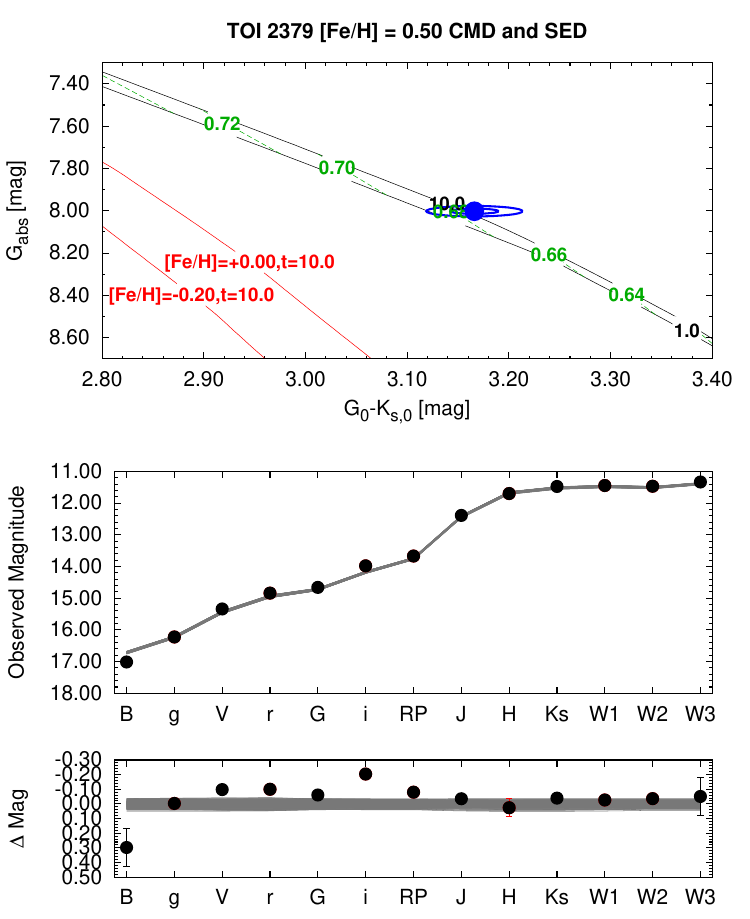}
    }
    \caption{Radial-velocity and SED  data for \Astartoi. \textbf{Left-top:} The radial-velocity data from ESPRESSO (circles) and PFS (squares) plotted with the systemic velocity estimated for each instrument subtracted and phase folded at the best-fit period from the modelling. The best-fit model is plotted as the solid line. The error-bars shown incorporate the jitter estimated for each instrument during the modelling. \textbf{Left middle:} The radial-velocity residuals from the fit. \textbf{Left bottom:} The bisector spans of the CCFs used to extract the radial-velocity measurements from the spectra.  \textbf{Right top:} Here we plot a colour-magnitude diagram, comparing the absolute \textit{Gaia G} magnitude to the dereddened $G - K_S$ colour, using magnitudes from \textit{Gaia} DR2 and 2MASS, shown as the blue circle, along with the 1$\sigma$ and 2$\sigma$ confidence level contours, shown as the blue lines. The solid black lines show theoretical isochrones with the ages in Gyr labelled, and the green dashed lines show MIST stellar evolution tracks interpolated using the best-fit value for the metallicity of the host star. The numerical labels for these tracks give the stellar mass (in units of \msun) for each evolution track.  \textbf{Right bottom: } the spectral energy distribution (SED) for the host star using observed magnitudes from broadband photometry, with the residuals to the SED modelling in the panel below. For both panels 200 randomly selected SED models from the MCMC posteriors are overplotted as the grey lines. }
    \label{fig:toi2379_rv_SED}
\end{figure*}

\begin{figure*}
    {
    \centering
    \includegraphics[width=0.49\linewidth]{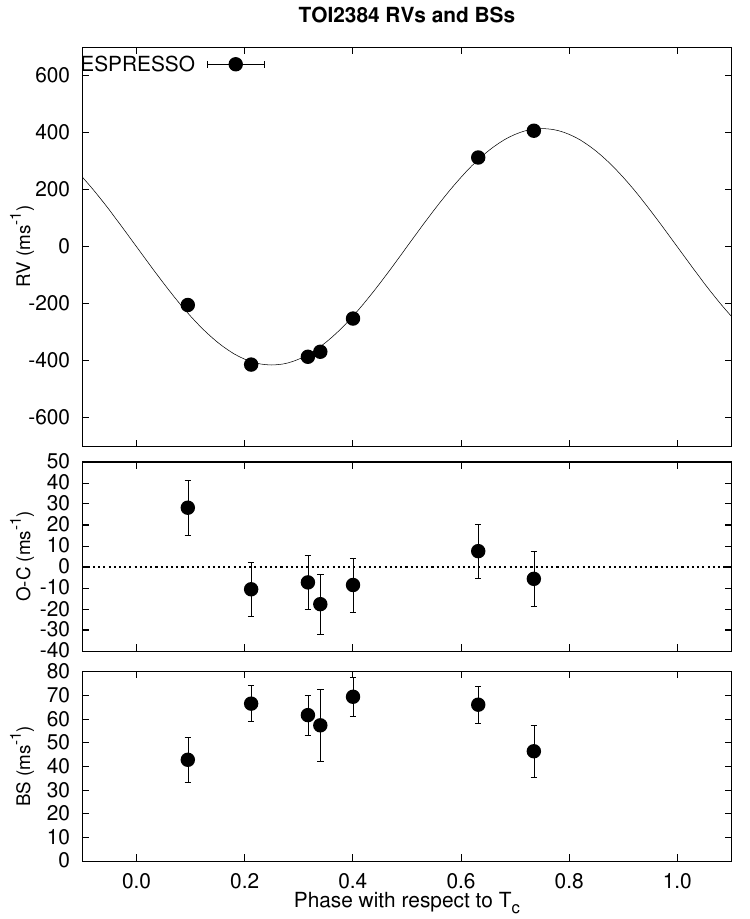}
    \hfill
    \includegraphics[width=0.49\linewidth]{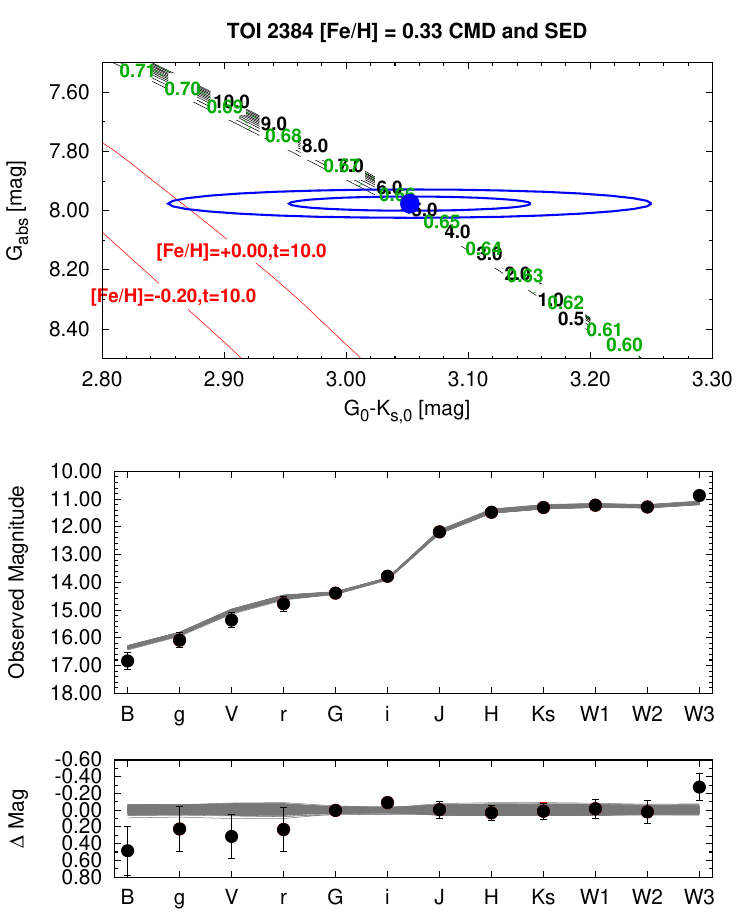}
    }
    \caption{Radial-velocity and SED  data for \Bstartoi. The panels and layout are the same as for Figure~\ref{fig:toi2379_rv_SED}.}
    \label{fig:toi2384_rv_SED}
\end{figure*}

We observed \Astartoi\ with the Planet Finder Spectrograph \citep[PFS;][]{crane2006pfs, crane2008pfs, crane:2010} on the Magellan Clay 6.5\,m telescope at Las Campanas Observatory in Chile. Observations were made on four nights between the UT dates of 2021 September 14 and 2021 November 16. A total of 10 exposures were obtained across the four nights through an I$_2$ cell, while five exposures were obtained without the I$_2$ cell on a single night. We used an exposure time of $20$\,minutes and read-out the data using 3$\times$3 binning mode. The I$_2$-free observations were combined to form a spectral template for precise relative RV measurements. We obtained RV measurements from the observations following \citet{butler:1996}, and spectral line bisector spans following a method similar to that described by \citet{torres:2007:hat3}. The RV and bisector span measurements are listed in Table~\ref{tab:toi2379_rv}, and are shown in Figure~\ref{fig:toi2379_rv_SED}.

\begin{table*}
    \centering
    \begin{tabular}{|c|c|c|c|c|c|}
    \hline
    \textbf{Time} & \textbf{Radial Velocity} & \textbf{Error} & \textbf{Bisector Span} & \textbf{Error} & \textbf{Instrument} \\
    BJD TDB & \ms & \ms & \ms & \ms & \\
    \hline
    2459497.60994212 & 27801.60 & 9.47 & 92.12 & 18.94 & ESPRESSO \\
    2459500.61738876 & 26881.81 & 5.37 & 58.10 & 10.75 & ESPRESSO \\
    2459519.61671364 & 27865.79 & 5.75 & 63.60 & 11.50 & ESPRESSO \\
    2459520.58923126 & 28547.72 & 3.77 & 35.62 & 7.53 & ESPRESSO \\
    2459521.63005011 & 28423.49 & 3.03 & 44.97 & 6.05 & ESPRESSO \\
    2459527.68871351 & 27157.53 & 3.16 & 39.45 & 6.33 & ESPRESSO \\
    2459528.59646265 & 26873.69 & 3.32 & 31.27 & 6.65 & ESPRESSO \\
    $ 2459471.72689 $ & $   162.90 $ & $    12.04 $ & \nodata      & \nodata      &  PFS \\
    $ 2459471.74115 $ & $    94.83 $ & $    13.99 $ & $ -1062.3 $ & $  899.5 $ &  PFS \\
    $ 2459471.75665 $ & $   223.48 $ & $    15.28 $ & $ -115.8 $ & $   91.0 $ &  PFS \\
    $ 2459475.72037 $ & $  -332.64 $ & $    26.50 $ & \nodata      & \nodata      &  PFS \\
    $ 2459531.55793 $ & \nodata      & \nodata      & $ -672.4 $ & $  297.7 $ &  PFS \\
    $ 2459531.57479 $ & \nodata      & \nodata      & $  -76.8 $ & $  407.8 $ &  PFS \\
    $ 2459531.59165 $ & \nodata      & \nodata      & $ -688.6 $ & $  583.5 $ &  PFS \\
    $ 2459531.60897 $ & \nodata      & \nodata      & $ -232.5 $ & $  280.6 $ &  PFS \\
    $ 2459531.62566 $ & \nodata      & \nodata      & $ -241.7 $ & $  258.5 $ &  PFS \\
    $ 2459531.64196 $ & $    -8.95 $ & $    11.86 $ & $ -705.5 $ & $  432.2 $ &  PFS \\
    $ 2459531.65633 $ & $    56.76 $ & $    12.61 $ & $  535.3 $ & $ 1170.9 $ &  PFS \\
    $ 2459531.67043 $ & $     0.00 $ & $    12.18 $ & $  -76.1 $ & $  488.8 $ &  PFS \\
    $ 2459534.55251 $ & $ -1476.56 $ & $    11.39 $ & $ -2000.8 $ & $  981.5 $ &  PFS \\
    $ 2459534.56710 $ & $ -1467.28 $ & $    10.24 $ & $  687.0 $ & $  357.8 $ &  PFS \\
    $ 2459534.58025 $ & $ -1424.81 $ & $    10.92 $ & $ 1771.0 $ & $  350.8 $ &  PFS \\
        \hline
    \end{tabular}
    \caption{Radial Velocities of \Astartoi. PFS observations without a radial velocity measurement correspond to I$_2$-free template observations, while those without a bisector span measurement had too low S/N to measure this quantity.}
    \label{tab:toi2379_rv}
\end{table*}

\begin{table*}
    \centering
    \begin{tabular}{|c|c|c|c|c|c|}
    \hline
    \textbf{Time} & \textbf{Radial Velocity} & \textbf{Error} & \textbf{Bisector Span} & \textbf{Error} & \textbf{Instrument} \\
    BJD TDB & \ms & \ms & \ms & \ms & \\
    \hline
    2459528.77980779 & 17808.12 & 4.13 & 69.47 & 8.27 & ESPRESSO \\
    2459530.78609980 & 17691.55 & 7.66 & 57.39 & 15.32 & ESPRESSO \\
    2459545.68721341 & 17673.80 & 4.23 & 61.70 & 8.46 & ESPRESSO \\
    2459548.71543659 & 18467.09 & 5.40 & 46.43 & 10.79 & ESPRESSO \\
    2459549.73434962 & 17646.34 & 3.80 & 66.57 & 7.60 & ESPRESSO \\
    2459552.76727365 & 18373.23 & 3.88 & 66.12 & 7.77 & ESPRESSO \\
    2459553.75552046 & 17855.60 & 4.85 & 42.81 & 9.70 & ESPRESSO \\
    \hline
    \end{tabular}
    \caption{Radial Velocities of \Bstartoi.}
    \label{tab:toi2384_rv}
\end{table*}

\section{Analysis}\label{sec:analysis}

\begin{table}
    \centering
    \caption{Stellar Properties for \Astartoi}
    \begin{tabular}{lcc} 
	\hline
	\multicolumn{3}{l}{\textbf{Identifiers}}\\
	\Astartoi \\
	\Astartic \\
	\textit{Gaia} DR2\,\AGAIAid \\
	\Atwomass \\
    \hline
    \textbf{Property}	&	\textbf{Value}		&\textbf{Source}\\
    \hline
    \multicolumn{3}{l}{\textbf{Astrometric Properties}}\\
    R.A.		&	\ARA			&{\em Gaia}	DR2\\
	Dec.		&	\ADec			&{\em Gaia}	DR2\\
    $\mu_{{\rm R.A.}}$ (\masy) & \ApropRA & {\em Gaia} DR2\\
	$\mu_{{\rm Dec.}}$ (\masy) & \ApropDec & {\em Gaia} DR2\\
	Parallax (mas)   &   \Aplx           &{\em Gaia} DR2\\
    \\
    \multicolumn{3}{l}{\textbf{Photometric Properties}}\\
	TESS (mag)  &\ATESSmag     &TIC8\\
    V (mag)		&\AVmag 	&APASS\\
	B (mag)		&\ABmag		&APASS\\
    \textit{Gaia} g (mag)&\AGAIAmag	&{\em Gaia} DR2\\
    \textit{Gaia} B$_P$ (mag)  &\AGAIABpmag    &{\em Gaia} DR2\\
    \textit{Gaia} R$_P$ (mag)  &\AGAIARpmag    &{\em Gaia} DR2\\
    J (mag)		&\AJmag		&2MASS	\\
   	H (mag)		&\AHmag		&2MASS	\\
	K (mag)		&\AKmag		&2MASS	\\
    WISE W1 (mag)	&\AWmag		&WISE	\\
    WISE W2 (mag)	&\AWWmag	&WISE	\\
    WISE W3 (mag)	&\AWWWmag		&WISE	\\
    \\
    \multicolumn{3}{l}{\textbf{Derived Properties}}\\
    \teff\ (K)    & \Ateffeccen               &Sec.~\ref{sec:analysis}\\
    \feh\	     & \Ametaleccen			    &Sec.~\ref{sec:analysis}\\
    \logg               & \Aloggeccen			&Sec.~\ref{sec:analysis}\\
    \mstar (\msun) & \Astarmasseccen	        &Sec.~\ref{sec:analysis}\\
    \rstar (\rsun) & \Astarradiuseccen	            &Sec.~\ref{sec:analysis}\\
    $\rho_*$ (\gccc) & \Astardensityeccen              & Sec.~\ref{sec:analysis}\\
    Age	(Gyr)			& \Aage				        	&Sec.~\ref{sec:analysis}\\
    Distance (pc)	&  \Adist	                &Sec.~\ref{sec:analysis}\\
	\hline
    \multicolumn{3}{l}{2MASS \citep{2MASS}; APASS \citep{APASS};}\\
    \multicolumn{3}{l}{{\em Gaia} DR2 \citep{GAIA_DR2}; TIC8 \citep{stassun2019tic8};}\\
    \multicolumn{3}{l}{WISE \citep{WISE}}\\
\end{tabular}
    \label{tab:toi2379_stellar}
\end{table}

\begin{table}
    \centering
    \caption{Stellar Properties for \Bstartoi}
    \begin{tabular}{lcc} 
	\hline
	\multicolumn{3}{l}{\textbf{Identifiers}}\\
	\Bstartoi \\
	\Bstartic \\
	\textit{Gaia} DR2\,\BGAIAid \\
	\Btwomass \\
    \hline
    \textbf{Property}	&	\textbf{Value}		&\textbf{Source}\\
    \hline
    \multicolumn{3}{l}{\textbf{Astrometric Properties}}\\
    R.A.		&	\BRA			&{\em Gaia}	DR2\\
	Dec.		&	\BDec			&{\em Gaia}	DR2\\
    $\mu_{{\rm R.A.}}$ (\masy) & \BpropRA & {\em Gaia} DR2\\
	$\mu_{{\rm Dec.}}$ (\masy) & \BpropDec & {\em Gaia} DR2\\
	Parallax (mas)   &   \Bplx           &{\em Gaia} DR2\\
    \\
    \multicolumn{3}{l}{\textbf{Photometric Properties}}\\
	TESS (mag)  &\BTESSmag     &TIC8\\
    V (mag)     &\BVmagorig      &APASS\\
    V (mag) - unblended		&\BVmag 	&Sec.~\ref{sec:analysis}\\
	B (mag)		&\BBmagorig		&APASS\\
    B (mag) - unblended		&\BBmag 	&Sec.~\ref{sec:analysis}\\
    \textit{Gaia} g (mag)&\BGAIAmag	&{\em Gaia} DR2\\
    \textit{Gaia} B$_P$ (mag)  &\BGAIABpmag    &{\em Gaia} DR2\\
    \textit{Gaia} R$_P$ (mag)  &\BGAIARpmag    &{\em Gaia} DR2\\
    J (mag)		&\BJmagorig		&2MASS	\\
    J (mag) - unblended		&\BJmag 	&Sec.~\ref{sec:analysis}\\
   	H (mag)		&\BHmagorig		&2MASS	\\
    H (mag) - unblended		&\BHmag 	&Sec.~\ref{sec:analysis}\\
	K (mag)		&\BKmagorig		&2MASS	\\
    K (mag) - unblended		&\BKmag 	&Sec.~\ref{sec:analysis}\\
    W1 (mag)	&\BWmagorig		&WISE	\\
    W1 (mag) - unblended	&\BWmag		&WISE	\\
    W2 (mag)	&\BWWmagorig	&WISE	\\
    W2 (mag) - unblended	&\BWWmag	&WISE	\\
    W3 (mag)	&\BWWWmagorig	&WISE	\\
    W3 (mag) - unblended	&\BWWWmag	&WISE	\\
    \\
    \multicolumn{3}{l}{\textbf{Derived Properties}}\\
    \teff\ (K)    & \Bteff               &Sec.~\ref{sec:analysis}\\
    \feh\	     & \Bmetal			    &Sec.~\ref{sec:analysis}\\
    \logg               & \Blogg			&Sec.~\ref{sec:analysis}\\
    \mstar (\msun) & \Bstarmass	        &Sec.~\ref{sec:analysis}\\
    \rstar (\rsun) & \Bstarradius	            &Sec.~\ref{sec:analysis}\\
    $\rho_*$ (\gccc) & \Bstardensity              & Sec.~\ref{sec:analysis}\\
    Age	(Gyr)			& \Bage				        	&Sec.~\ref{sec:analysis}\\
    Distance (pc)	&  \Bdist	                &Sec.~\ref{sec:analysis}\\
	\hline
    \multicolumn{3}{l}{2MASS \citep{2MASS}; APASS \citep{APASS};}\\
    \multicolumn{3}{l}{{\em Gaia} DR2 \citep{GAIA_DR2}; TIC8 \citep{stassun2019tic8};}\\
    \multicolumn{3}{l}{WISE \citep{WISE}}\\
\end{tabular}
    \label{tab:toi2384_stellar}
\end{table}

We performed a joint analysis of all available data, including broadband photometry, transit discovery and follow-up light curves, astrometric data, and radial velocity measurements, in order to derive and constrain the stellar and planetary parameters for the \Astartoi\ and \Bstartoi\ systems. The analysis followed the methods of \citet{hartman2019hats60_69} and \citet{bakos2020hats71}. In short, the transit light curves were modelled using \citet{mandelagol2002transitmodel} models with limb-darkening coefficients constrained using priors derived for the theoretical models from \citet{claret2012LD, claret2013LD, claret2018LD} and the RV measurements were modelled assuming a Keplerian orbit for the exoplanets. In order to constrain the stellar parameters, we include in the fit the broad-band photometry from \textit{Gaia}, 2MASS, APASS and WISE (listed in Tables~\ref{tab:toi2379_stellar}~and~\ref{tab:toi2384_stellar}), \textit{Gaia} DR3 parallax (listed in Tables~\ref{tab:toi2379_stellar}~and~\ref{tab:toi2384_stellar}), and spectroscopically derived stellar atmospheric parameters. The physical parameters of the stars are forced to be consistent with the MIST version 1.2 stellar evolution models \citep{paxton:2011,paxton:2013,paxton:2015,choi:2016,dotter:2016}, with this constraint incorporated directly into the joint modelling of the data. We accounted for systematic errors in the evolution models following the method described by \citet{hartman:2023}. We used the MWDUST 3D Galactic extinction model \citep{bovy:2016} to place a Gaussian prior on the line-of-sight extinction $A_{V}$ vs.\ distance, and to set a maximum value for this parameter.

For \Bstartoi\ we correct all of the catalog photometry and light curves for blending from the $\sim 0\farcs9$ neighbor listed in the \textit{Gaia}~DR3 catalog. Here we assume the neighbor is a physical companion to \Bstartoi\ (the similar parallax and proper motion for the neighbor compared to the planet hosting star supports this assumption), and use the difference in the absolute $G$-band magnitude from \Bstartoi, together with the MIST evolution models, to determine the physical properties of the neighbor, which we find to be a $\sim 0.44$\,\msun\ star, and to estimate its expected brightness in each band-pass. The estimated flux from the neighbor is then subtracted from the catalog magnitudes for \Bstartoi\ that are included in the analysis of the system. The original and corrected values are listed for several band-passes in Table~\ref{tab:toi2384_stellar}. Here the uncertainties in the properties of the neighbor are propagated into the correction for blending, which causes the unblended magnitudes to have larger uncertainties than the original magnitudes. We assume that the \textit{Gaia} magnitudes are not blended. For the light curves we allow blending factors to vary in the fit as free parameters, with priors set to the expected flux contributions in each band-pass from the neighbor.

We fit the observations using a differential evolution Markov Chain Monte Carlo procedure, assuming priors on the parameters as listed in \citet{hartman2019hats60_69}. To assess the convergence of the MCMC chains we computed the integrated correlation time for all the parameters following \citet{goodmanweare2010} and ran the sampling until the total chain length, after removing the burn-in phase, exceeded 20 correlation lengths for all parameters. We then calculate the median and $1\sigma$ uncertainty bounds for each parameter from the a posteriori distributions.

We used the \textsc{ODUSSEAS} software \citep{ODUSSEAS2020} to derive \feh\ and \teff\ values for the host star from the ESPRESSO spectra. \textsc{ODUSSEAS} is a machine learning based code designed specifically to perform spectral analysis for M-dwarf stars \citep[e.g.][]{lillobox2020lhs1140,Hobson2023toi3235}. Using \textsc{ODUSSEAS} we derive values of \feh$ = $\AmetalODUSSEAS\ and \teff$ = $\AteffODUSSEAS\,K for \Astartoi\ and \feh$ = $\BmetalODUSSEAS\ and \teff$ = $\BteffODUSSEAS\,K for \Bstartoi, which we adopt as priors for the joint analysis.

For both systems, we performed one analysis assuming a circular orbit, and one allowing for an eccentric orbit. Allowing the eccentricity of \Bstartoi\ to be fit as a free parameter we find a value of $e = 0.012 \pm 0.014$, with a 2$\sigma$ upper limit of $e < 0.043$. This result is fully consistent with a circular orbit, and so we conclude we find no evidence of an eccentric orbit and adopt the model with the fixed circular orbit as the best-fit model for this system. For \Astartoi\ the radial-velocity measurements favour an eccentric orbit with an eccentricity of \Aecc. Comparing the eccentric orbit model with the circular orbit model, we find a Bayesian Information Criterion difference $\Delta\textrm{BIC} = - 252.39$, which represents very strong evidence in favour of the eccentric orbit. Therefore, we adopt the model with the eccentric orbit as the best-fit model for \Aplanettoi.

\subsection{Blend scenarios}\label{sub:blendscenarios}
For both systems presented in this work, we perform additional modelling to investigate the possibility that the photometric and spectroscopic signals observed are a result of a blended stellar binary or triple system. This blend analysis follows the procedure presented in \citet{hartman2019hats60_69} and \citet{bakos2020hats71}. In short, four different scenarios are considered: a single star with a transiting planet, a wide binary star system with a planet transiting one of the stars, a hierarchical triple system formed of a bright star and two fainter stars in an eclipsing binary pair, and a bright foreground star with a background eclipsing binary.

For \Bstartoi\ the single star with a transiting planet model provides the best fit to the data. The best fit blend scenario is for the background eclipsing binary scenario, although this model has $\Delta \chi^2 = + 17.3$ compared to the single star model. As such, we can confidently rule out any blend scenarios for \Bstartoi.

For \Astartoi, the best fit blended stellar eclipsing binary model yields a $\chi^2$ value indistinguishable from the best fit model for a single star with a transiting planet. Computing the Bayesian Information Criterion (BIC), which takes the number of free parameters into account, we find a significantly lower BIC for the single star and planet model ($\Delta\textrm{BIC} = -45.9$) compared to the best fit blend model. We also note that any blend models predict bisector span variations significantly larger than that seen in the ESPRESSO observations. Therefore, we confidently conclude that the \Astartoi\ system consists of a planet transiting a single star. 

\begin{table*}
    \centering
    \caption{Planet parameters for \Aplanettoi\ and \Bplanettoi}
    \begin{tabular}{lcccc}
    \multicolumn{3}{c}{\textbf{Identifiers}} & \multicolumn{2}{c}{Values} \\
    \noalign{\smallskip} 
    Name & Symbol & Unit & \Aplanettoi & \Bplanettoi \\
    \noalign{\smallskip} 
    \hline
    \noalign{\smallskip} 
    Transit Mid-Point Time & \tc & BJD (TDB) & \Atc & \Btc \\
    \noalign{\smallskip} 
    Orbital Period & $P$ & days & \Aperiod & \Bperiod \\
    \noalign{\smallskip} 
    Radius Ratio & \rprs & & \Arratio & \Brratio \\
    \noalign{\smallskip} 
    Scaled Semi-major Axis & $a / R_*$ & & \Aaoverr & \Baoverr \\
    \noalign{\smallskip} 
    Impact Parameter & $b$ & & \Aimpact & \Bimpact \\
    \noalign{\smallskip} 
    Orbital Inclination & $i$ & degrees & \Ainc & \Binc \\
    \noalign{\smallskip} 
    Transit Duration & $T_{\rm dur}$ & hours & \Aduration & \Bduration \\
    \noalign{\smallskip} 
    RV Semi-Amplitude & $K$ & \ms & \Asemiamp & \Bsemiamp \\
    \noalign{\smallskip} 
    Orbital Eccentricity & $e$ & & \Aecc & 0. (fixed)\\
    \noalign{\smallskip} 
    Argument of Pericentre & $\omega$ & degrees & \Aomega & \nodata \\
    \noalign{\smallskip} 
    Planet Radius & \rpl & \rjup & \Aradius & \Bradius \\
    \noalign{\smallskip} 
    Planet Mass & \mpl & \mjup & \Amass & \Bmass \\
    \noalign{\smallskip} 
    Planet-to-Star Mass Ratio & \mpl / \mstar & \% & \Amassratio & \Bmassratio \\
    \noalign{\smallskip} 
    Semi-major Axis & $a$ & AU & \Aau & \Bau \\
    \noalign{\smallskip} 
    Planet Density & \rhopl & \gccc & \Adensity & \Bdensity \\
    \noalign{\smallskip}
    Planet Equilibrium Temperature$^{A}$ & $T_{\rm eq}$ & $K$ & \Ateqave & \Bteq \\
    \noalign{\smallskip}
    Planet Irradiation Flux$^{A}$ & $F_{\rm irrad}$ & erg\,cm$^{-2}$\,s$^{-1}$ & \Ainsolave & \Binsol \\
    \noalign{\smallskip}
    \hline
    \hline
    \noalign{\smallskip} 
    ESPRESSO Systemic RV & $\Gamma_{\rm RV; ESPRESSO}$ & \ms & \AgammaESP & \BgammaESP \\
    \noalign{\smallskip} 
    PFS Systemic RV & $\Gamma_{\rm RV; PFS}$ & \ms & \AgammaPFS & \nodata \\
    \noalign{\smallskip} 
    ESPRESSO RV Jitter & $\sigma_{\rm RV; ESPRESSO}$ & \ms & \AjitterESP & \BjitterESP \\
    \noalign{\smallskip} 
    PFS RV Jitter & $\sigma_{\rm RV; PFS}$ & \ms & \AjitterPFS & \nodata \\
    \noalign{\smallskip} 
    \hline    
    \multicolumn{5}{l}{A - for \Aplanettoi\ we provide the orbit averaged values.} \\
\end{tabular}
    \label{tab:planet_params}
\end{table*}

\section{Discussion}\label{sec:discuss}
We find both \Astartoi\ and \Bstartoi\ to host transiting giant planets, deriving masses and radii of \mpl$ = $\Amass\,\mjup\ and \rpl$ = $\Aradius\,\rjup\ for \Aplanettoi\ and \mpl$ = $\Bmass\,\mjup\ and \rpl$ = $\Bradius\,\rjup\ for \Bplanettoi. We also constrain the two host stars to have masses of \mstar$ = $\Astarmasseccen\,\msun\ for \Astartoi\ and \mstar$ = $\Bstarmass\,\msun\ for \Bstartoi. We provide the full results for the host star parameters in Tables~\ref{tab:toi2379_stellar}~\&~\ref{tab:toi2384_stellar} and the full results for the planet parameters in Table~\ref{tab:planet_params}.

\subsection{\Aplanettoi\ and \Bplanettoi\ in context}\label{sub:context}
\begin{figure*}
    \centering
    \includegraphics[width=0.48\linewidth]{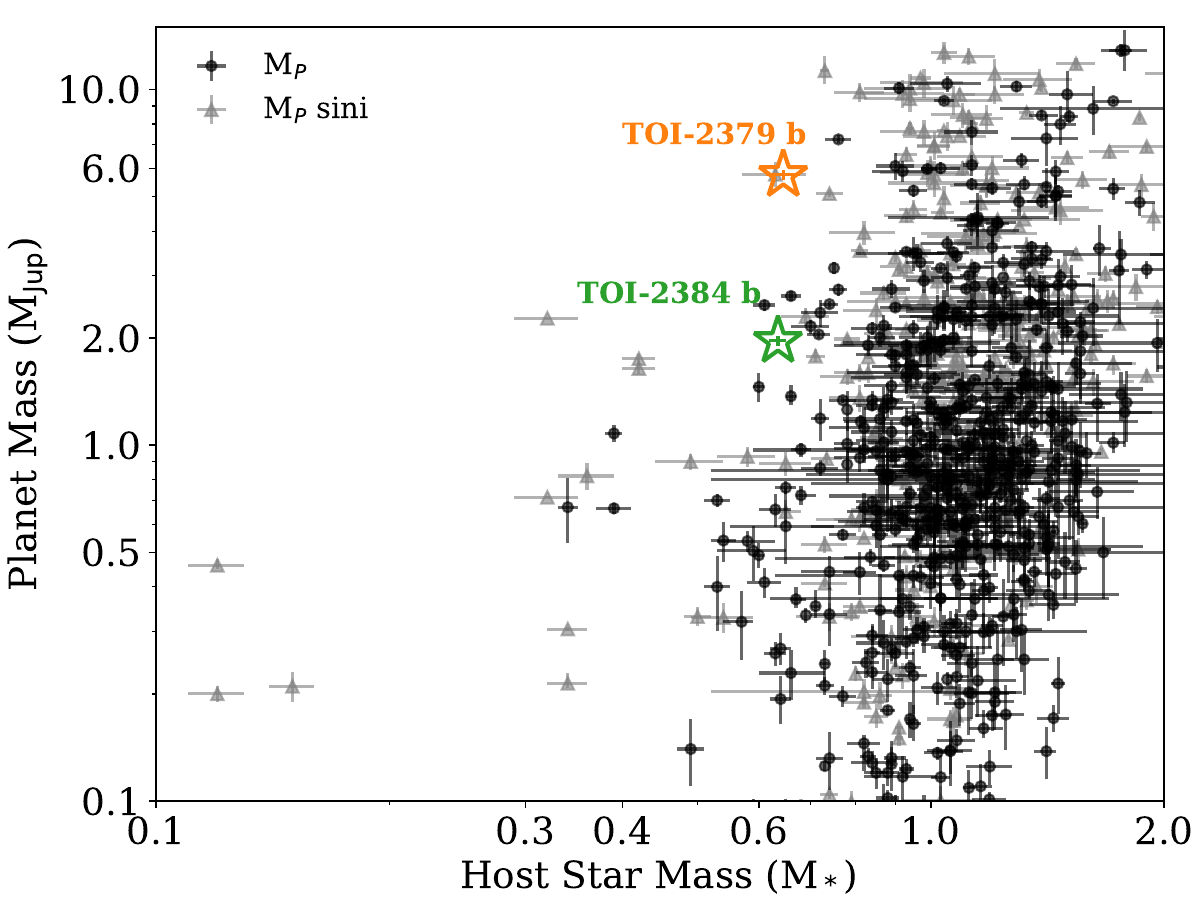}
    \hfill
    \includegraphics[width=0.48\linewidth]{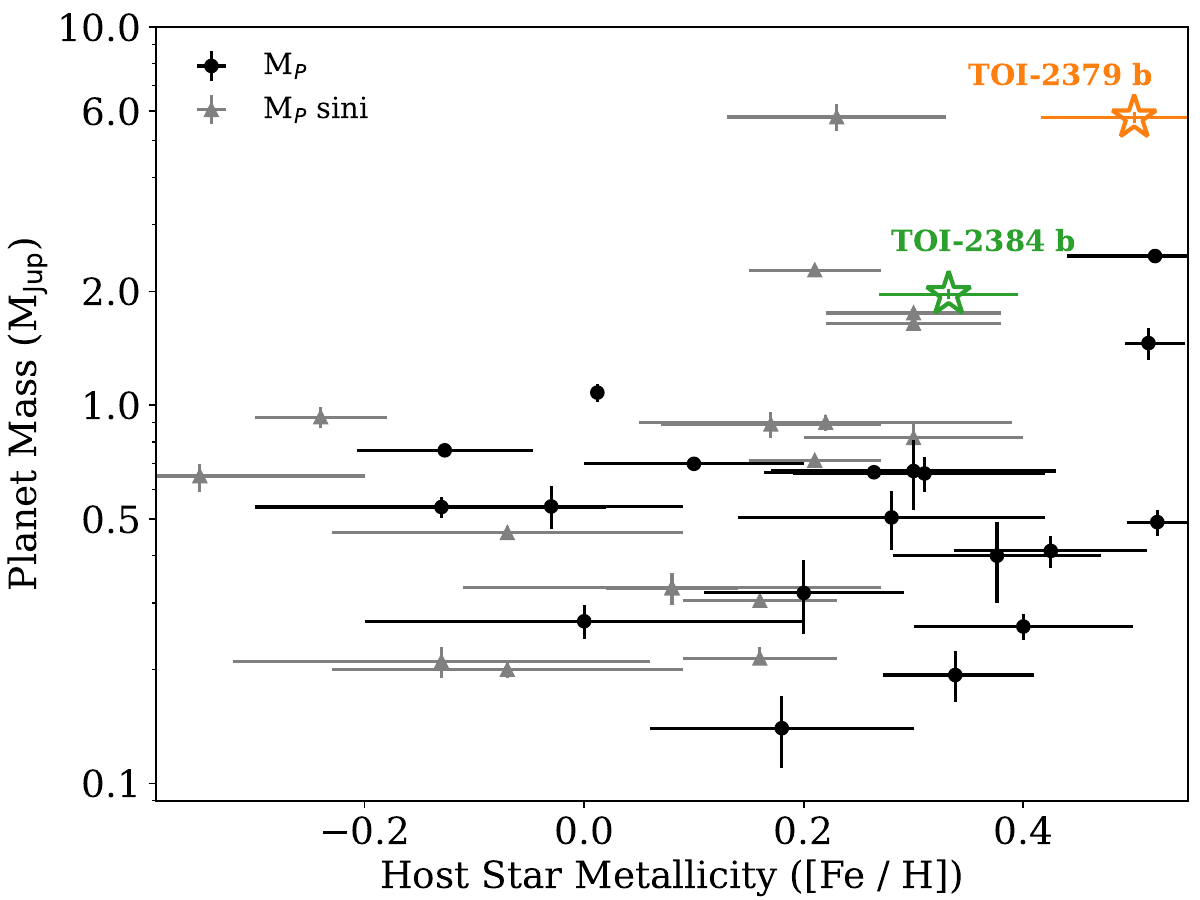}
    \caption{\Aplanettoi\ and \Bplanettoi\ in the context of known giant exoplanets ($\mpl \geq 0.1\mjup$). In both plots \Aplanettoi\ is plotted as the orange star and \Bplanettoi\ as the green star. \textbf{Left:} The population of known exoplanets plotted comparing the mass of the planet to the mass of the host star. The known exoplanet population plotted has been drawn from exoplanets discovered through the transit and radial velocity methods which have a planet mass measured to better then 10\,\% precision. The black circles give the known planets with a measured absolute mass and a radius measured to better than 30\,\% precision. The grey triangles denote the radial velocity planets which do not transit and do not have a known radius for which we plot the minimum mass $\mpl\sin i$. We accessed the data from the NASA exoplanet archive, accessed on 2nd February 2024. \textbf{Right:} The population of known giant exoplanets with low-mass host stars ($\mstar \leq 0.65\msun$) plotted comparing the planet mass with the metallicity of the host star. The selection criteria and markers used are the same as for the left-hand plot.}
    \label{fig:mstar_vs_mpl}
\end{figure*}
We compare \Aplanettoi\ and \Bplanettoi\ to the population of known exoplanets in Figure~\ref{fig:mstar_vs_mpl} finding \Aplanettoi\ and \Bplanettoi\ to be the most and third most massive transiting exoplanets with a low-mass host star ($\mstar \leq 0.65\,\msun$). Considering the planet-to-star mass ratios of these systems we find a mass ratio of \Amassratio\,\% for \Aplanettoi, the highest of any transiting exoplanet with a low-mass host star and the second highest mass ratio of all exoplanets with low-mass hosts, second only to GJ~676~A~b \citep{forveille2011gj676HARPS,sahlmann2016gj676Astrom}. This mass ratio is also the fifth highest for transiting exoplanets across all stellar masses. The planet-to-star mass ratio for \Bplanettoi\ is \Bmassratio\,\%, the third highest out of transiting exoplanets with low-mass host stars behind \Aplanettoi\ and TOI-4201~b \citep{gan2023toi4201}. These high mass ratios make these two systems extremely intriguing from a viewpoint of understanding the extremes of how giant planets can form.

What is equally interesting to observe from Figure~\ref{fig:mstar_vs_mpl} is the region of parameter space with $\mstar \leq 0.6\msun$ and $\mpl \gtrsim 2\mjup$ which is bare of planets. This lack of very massive planets for low-mass stars is as expected from core-accretion formation theory \citep{burn2021ngppslowmassstars}. It also lies in contrast to the population of planets with masses similar to Saturn and Jupiter ($0.2 \mjup < \mpl < 1.0 \mjup$) which extends down to host stars as low-mass as 0.3\,\msun, with a further handful of roughly Saturn mass planets with host stars as low-mass as 0.1\,\msun. It is this population of giant planets for $\mstar \leq 0.4 \msun$ host stars which poses the largest challenge for core-accretion \citep[e.g.][]{burn2021ngppslowmassstars, Hobson2023toi3235}. 

The planets we present in this work on the other hand, while pushing the boundaries of the population of known planets, can be reconciled with core-accretion formation theory. With masses between $0.6 - 0.65 \msun$, the population synthesis predictions of \citet{burn2021ngppslowmassstars} show that giant planets could form around these host stars. Moreover, \citet{burn2021ngppslowmassstars} do not predict a decrease in the masses of the giant planets that form around these low-mass host stars, compared to giant planets with Solar-like host stars. In fact, for their simulations with a 0.7\,\msun\ host star -- the closest in mass to the host stars of our two new planets -- they predict the presence of a population of giant planets ranging in mass from approximately 1\,\mjup\ up to 12\,\mjup. \Aplanettoi\ and \Bplanettoi, with masses of \Amass\,\mjup\ and \Bmass\,\mjup\ respectively, sit well within this range. Therefore, following the predictions of \citet{burn2021ngppslowmassstars}, it is reasonable that these two planets could have formed through core-accretion.

The formation of giant planets through core-accretion for host stars of any mass has been linked to a high metallicity of the host stars \citep[e.g.][]{fischer2005pmc,johnson2010gplmetallicity,sousa2011harpsPMC,osborn2020pmc}. In Figure~\ref{fig:mstar_vs_mpl} we plot the masses for known giant planets with low-mass host stars ($\mstar \leq 0.65 \msun$) as a function of the metallicity of the hosts. While we find that giant planets with masses $\mpl \lesssim 1.0 \mjup$ can exist for low-mass stars of a wide range of sub- and super-solar metallicities, more massive giant planets ($\mpl > 1.1 \mjup$) only exist for host stars with very high metallicities $\feh \geq 0.2$\,dex. The restriction of super-Jupiter planets to high metallicity stars again implies a core-accretion formation process for these stars \citep[e.g.][]{idalin2004metallicityformation,emsenhuber2021ngpps2}, as the highly metal-enriched protoplanetary disks around these stars would provide a more favourable location for the formation of massive planets. 

\subsection{Orbital Eccentricity of \Aplanettoi}\label{sub:toi2379_eccentricity}
From our analysis we determined the orbit of \Aplanettoi\ to be significantly eccentric with a measured eccentricity of \Aecc. It is expected that the orbits of close-orbiting planets will circularize over time due to the tidal interactions between the star and planet \citep[e.g.][]{rasioford1996tidalcircularization}. Using the equations from \citet{adams2006tidalcircularization} we can estimate the circularization timescale, $\tau_{\rm circ}$, for \Aplanettoi\ to determine whether we would have expected the orbit to have circularized. The circularization timescale depends strongly on a quantity known as the tidal quality factor $Q_P$. Estimating the value of $Q_P$ for a single planet is very difficult as there exist a wide number of different proposed theoretical models for the tidal dissipation in planets \citep[see][and the references therein]{mahmud2023tidalcircularization}. Various efforts to empirically constrain $Q_P$ using the known population of hot Jupiters have been performed and have obtained differing results for $Q_P$. 

For this analysis we consider the results of three studies: $\log_{10} Q_P \sim 6.5$ \citep{jackson2008qp}, $\log_{10} Q_P = 6.14^{+0.41}_{-0.25}$ \citep{quinn2014qp}, and $\log_{10} Q_P = 5.0 \pm 0.5$ \citep{mahmud2023tidalcircularization}. Combining these three results, we use a range of $4.5 \leq \log_{10} Q_P \leq 6.5$ to estimate possible tidal circularization timescales for \Aplanettoi, obtaining $0.12 {\rm Gyr} \leq \tau_{\rm circ} \leq 16.6 {\rm Gyr}$. From our isochrone analysis (section~\ref{sec:analysis}) we estimate an age of the \Astartoi\ system of $13.8 \pm 4.1$\,Gyr, although from Figure~\ref{fig:toi2379_rv_SED} we can see that the age is consistent with being as low as 1\,Gyr to within a confidence of 2$\sigma$. Given the large uncertainty on both $Q_P$ and the age of the \Astartoi\ system, we are unable to confidently say whether we would have expected the orbit to have tidally circularized yet. In the scenario in which $\tau_{\rm circ}$ is significantly less than the age of the system some other interaction would be needed to be responsible for the eccentric orbit, such as with an outer massive body in the system. Further work into improving our knowledge of both the age of the system as well as $Q_P$ is required to determine whether the presence of such a companion is required to explain the orbit of \Aplanettoi. With long-term radial velocity monitoring of \Astartoi\ over a number of months and years we would be able to search for any such companions however with just a handful of observations spanning just two months at this stage we are unable to place any constraints on companions in the system. The existence of such a companion in the \Astartoi\ system could have interesting implications for the formation history of \Aplanettoi.

\subsection{Prospects for future follow-up}\label{sub:future_followup}
By studying the atmospheric compositions of \Aplanettoi\ and \Bplanettoi\ we could uncover further information into the formation processes and migration histories of these two planets \citep[e.g.][]{madhusudham2019atmospheresreview,hobbs2022formationatmospheres}. With transmission spectroscopy metrics \citep[TSM;][]{kempton2018tsm} of just 5.21 (\Aplanettoi) and 23.36 (\Bplanettoi) atmospheric characterization of these two planets will be  tough yet possible with JWST \citep{gardner2006jwst}. That said, as the two most massive transiting exoplanets with host stars less massive than 0.65\,\msun\ the characterisation of their atmospheres will be a crucial piece of the puzzle when it comes to fully understanding how these exotic systems form and evolve.

\section{Conclusions}
We report the discovery of \Aplanettoi\ and \Bplanettoi, two super-Jupiter mass giant planets with metal-rich low-mass host stars. We derive the planetary and stellar parameters from a joint analysis of transit photometry, high precision radial-velocity measurements, and broadband photometry, finding \Aplanettoi\ and \Bplanettoi\ to be the most and third most massive transiting exoplanets with a low-mass host star ($\mstar \leq 0.65\,\msun$). We also find the host stars for both planets to be very metal rich providing some further clues into the formation history of these exotic planetary systems.

Over the next few years, further discoveries and precise mass measurements of giant planets orbiting low-mass stars will help to reveal whether any of the trends we are beginning to see are maintained. The TESS mission is set to play a large role in this. Already nearly one hundred candidate giant planets with low-mass host stars have been found by TESS, primarily from searches using Full-Frame-Image data \citep[e.g.][]{kunimoto2022qlpfaintstar, bryant2023lmstargiantplanetoccrates}. Confirming and characterizing these and other candidate planets will allow us to study this exotic planet population in much better detail than currently possible.

\section*{Acknowledgements}

The contributions at the Mullard Space Science Laboratory by E.M.B. have been supported by STFC through the consolidated grant ST/W001136/1. 
The postdoctoral fellowship of KB is funded by F.R.S.-FNRS grant T.0109.20 and by the Francqui Foundation.
F.J.P acknowledges financial support from the grant CEX2021-001131-S
funded by MCIN/AEI/10.13039/501100011033 and through projects
PID2019-109522GB-C52 and PID2022-137241NB-C43.
J.D.H and G.A.B. acknowledge funding from NASA XRP grant 80NSSC22K0315.
This publication benefits from the support of the French Community of Belgium in the context of the FRIA Doctoral Grant awarded to MT

The research leading to these results has received funding from the ARC grant for Concerted Research Actions, financed by the Wallonia-Brussels Federation. TRAPPIST is funded by the Belgian Fund for Scientific Research (Fond National de la Recherche Scientifique, FNRS) under the grant PDR T.0120.21. MG is F.R.S.-FNRS Research Director and EJ is F.R.S.-FNRS Senior Research Associate. Observations were carried out from ESO La Silla Observatory.

Funding for the TESS mission is provided by NASA's Science Mission Directorate. KAC and CNW acknowledge support from the TESS mission via subaward s3449 from MIT.
This paper made use of data collected by the TESS mission, obtained from the Mikulski Archive for Space Telescopes MAST data archive at the Space Telescope Science Institute (STScI). Funding for the TESS mission is provided by the NASA Explorer Program. STScI is operated by the Association of Universities for Research in Astronomy, Inc., under NASA contract NAS 5–26555.
We acknowledge the use of public TESS data from pipelines at the TESS Science Office and at the TESS Science Processing Operations Center. 
Resources supporting this work were provided by the NASA High-End Computing (HEC) Program through the NASA Advanced Supercomputing (NAS) Division at Ames Research Center for the production of the SPOC data products.

We acknowledge funding from the European Research Council under the ERC Grant Agreement n. 337591-ExTrA.
This work makes use of observations from the LCOGT network. Part of the LCOGT telescope time was granted by NOIRLab through the Mid-Scale Innovations Program (MSIP). MSIP is funded by NSF.
This research has made use of the Exoplanet Follow-up Observation Program (ExoFOP; DOI: 10.26134/ExoFOP5) website, which is operated by the California Institute of Technology, under contract with the National Aeronautics and Space Administration under the Exoplanet Exploration Program.

\section*{Data Availability}

The TESS photometry used is all publicly available from the MAST data archive. The follow-up photometry (except for the ExTrA observations) is available from the ExoFoP web pages for each object. The ExTrA observations are available on request. The radial velocity data used are available from Tables~\ref{tab:toi2379_rv}~\&~\ref{tab:toi2384_rv} in this work.



\bibliographystyle{mnras}
\bibliography{paper} 





\appendix

\section{Affiliations}\label{sec:affil}
$^{1}$Mullard Space Science Laboratory, University College London, Holmbury St Mary, Dorking, RH5 6NT, UK\\
$^{2}$Department of Physics, University of Warwick, Gibbet Hill Road, Coventry CV4 7AL, UK\\
$^{3}$Centre for Exoplanets and Habitability, University of Warwick, Gibbet Hill Road, Coventry CV4 7AL, UK\\
$^{4}$Department of Astrophysical Sciences, Princeton University, Princeton, NJ 08544, USA\\
$^{5}$European Southern Observatory (ESO), Av. Alonso de Córdova 3107, 763 0355 Vitacura, Santiago, Chile\\
$^{6}$Max Planck Institute for Astronomy, K\"onigstuhl 17, D-69117—Heidelberg, Germany\\
$^{7}$Millennium Institute of Astrophysics (MAS), Nuncio Monseñor Sótero Sanz 100, Providencia, Santiago, Chile\\
$^{8}$Facultad de Ingenier\'ia y Ciencias, Universidad Adolfo Ib\'a\~nez, Av. Diagonal las Torres 2640, Pe\~nalol\'en, Santiago, Chile\\
$^{9}$Data Observatory Foundation, Santiago, Chile\\
$^{10}$Univ. Grenoble Alpes, CNRS, IPAG, F-38000 Grenoble, France\\
$^{11}$Observatoire de Gen\`eve, Département d’Astronomie, Universit\'e de Gen\`eve, Chemin Pegasi 51b, 1290 Versoix, Switzerland\\
$^{12}$Astrobiology Research Unit, Universit\'e de Li\`ege, 19C All\'ee du 6 Ao\^ut, 4000 Li\`ege, Belgium\\
$^{13}$Department of Earth, Atmospheric and Planetary Science, Massachusetts Institute of Technology, 77 Massachusetts Avenue, Cambridge, MA 02139, USA\\
$^{14}$Instituto de Astrof\'isica de Canarias (IAC), Calle V\'ia L\'actea s/n, 38200, La Laguna, Tenerife, Spain\\
$^{15}$Center for Astrophysics \textbar \ Harvard \& Smithsonian, 60 Garden Street, Cambridge, MA 02138, USA\\
$^{16}$School of Physics \& Astronomy, University of Birmingham, Edgbaston, Birmingham B15 2TT, United Kingdom\\
$^{17}$El Sauce Observatory, Coquimbo Province, Chile\\
$^{18}$ Space Sciences, Technologies and Astrophysics Research (STAR) Institute, Universit\'e de Li\`ege, All\'ee du 6 Ao\^ut 19C, B-4000 Li\`ege, Belgium\\
$^{19}$Departamento de Astrof\'isica, Universidad de La Laguna (ULL), E-38206 La Laguna, Tenerife, Spain\\
$^{20}$ Instituto de Astrofísica de Andalucía (IAA-CSIC), Glorieta de la Astronomía s/n, 18008 Granada, Spain\\
$^{21}$Observatoire des Baronnies Provençales (OBP), Laboratoire d'Astrophysique des Baronnies - Science Citoyenne et Actions pour la Nuit (LABSCAN),\\ 1480 Route des Alpes, 05150 Moydans, France\\
$^{22}$Earth and Planets Laboratory, Carnegie Institution for Science, 5241 Broad Branch Road, NW, Washington, DC 20015, USA\\
$^{23}$The Observatories of the Carnegie Institution for Science, 813 Santa Barbara Street, Pasadena, CA, 91101, USA\\
$^{24}$Department of Physics and Astronomy, The University of New Mexico, 210 Yale Blvd NE, Albuquerque, NM 87106, USA\\
$^{25}$NASA Ames Research Center, Moffett Field, CA 94035, USA\\
$^{26}$Space Telescope Science Institute, 3700 San Martin Drive, Baltimore, MD, 21218, USA\\
$^{27}$Department of Physics and Kavli Institute for Astrophysics and Space Research, Massachusetts Institute of Technology, Cambridge, MA 02139, USA\\
$^{28}$Department of Aeronautics and Astronautics, MIT, 77 Massachusetts Avenue, Cambridge, MA 02139, USA



\bsp	
\label{lastpage}
\end{document}